\documentclass[aps,prx,twocolumn,superscriptaddress,groupedaddress,longbibliography]{revtex4-2}
\usepackage{braket}
\usepackage{graphicx}
\usepackage{amsmath}    
\usepackage{amsfonts}
\usepackage{amsmath}
\usepackage{lipsum}
\usepackage{xcolor}
\usepackage{qcircuit}
\usepackage{mhchem}
\usepackage{amssymb}

\newcommand\normx[1]{\Vert#1\Vert}

\begin{document}
\title{A stochastic quantum Krylov protocol with double factorized Hamiltonians}

\author{Nicholas H. Stair}
\email{nick.stair@qcware.com}
\author{Cristian L. Cortes}
\email{cris.cortes@qcware.com}
\author{Robert M. Parrish}
\email{rob.parrish@qcware.com}
\affiliation{QC Ware Corporation, Palo Alto, California 94301, USA}
\author{Jeffrey Cohn}
\email{jeffrey.cohn@ibm.com}
\author{Mario Motta}
\email{mario.motta@ibm.com}
\affiliation{IBM Quantum, IBM Research – Almaden, San Jose, California 95120, USA}
\date{\today}

\begin{abstract}
We propose a class of randomized quantum Krylov diagonalization (rQKD) algorithms capable of solving the eigenstate estimation problem with modest quantum resource requirements.
Compared to previous real-time evolution quantum Krylov subspace methods, our approach expresses the time evolution operator, $e^{-i\hat{H} \tau}$, as a linear combination of unitaries and subsequently uses a stochastic sampling procedure to reduce circuit depth requirements. 
While our methodology applies to any Hamiltonian with fast-forwardable subcomponents, we focus on its application to the explicitly double-factorized electronic-structure Hamiltonian.
To demonstrate the potential of the proposed rQKD algorithm, we provide numerical benchmarks for a variety of molecular systems with circuit-based statevector simulators, achieving ground state energy errors of less than 1~kcal~mol$^{-1}$ with circuit depths orders of magnitude shallower than those required for low-rank deterministic Trotter-Suzuki decompositions. 
\end{abstract}

\maketitle

\section*{Introduction}

Efficient determination of eigenpairs for quantum many-body systems is one of the most important computational challenges in modern physics, chemistry, and materials science. 
Some of the most promising algorithms involve the use of quantum computers to circumvent the naively exponential classical storage complexity for many-body states. 
Such approaches often utilize the unitary evolution of the time-dependent Schr\"{o}dinger equation (deemed quantum simulation) as a cornerstone subroutine, a problem contained in the bounded-error quantum polynomial (BQP) complexity class~\cite{Georgescu:2014um}.
The most extensively studied quantum algorithms for eigenpair determination rely on an combination of quantum simulation and quantum phase estimation (QPE)~\cite{Abrams:1999ur}, but recent attention has been paid to a promising family of variational quantum simulation algorithms we will refer to as quantum Krylov diagonalization (QKD)~\cite{parrish2019quantum, stair:Krylov2020, Klymko:2022vp}. 
Presently, usage of QPE or QKD is limited by the substantial gate counts required for quantum simulation via standard Trotter-Suzuki decompositions \cite{SUZUKI1990319, Suzuki_1991}, an issue which has seen a chronology of improvements including the exploitation of Hamiltonian sparsity \cite{mcclean2014exploiting}, low-rank Hamiltonian factorization \cite{poulin2014trotter, Motta_2021, Kivlichan:2021ld, huggins2021efficient}, stochastic compilation methods \cite{childs2019faster, Campbell:2019qdrift, kivlichan2019phase, ouyang2020compilation, wan2022randomized}, and a variety of post-Trotter methods~\cite{childs:2012lcu, Babbush_2016, low:2017qsp, Berry2019qubitizationof}.  
Although significant progress has been made in reducing the general cost of quantum simulation for fault-tolerant hardware, there has been much less attention paid to improving quantum simulation specifically designed for QKD-type algorithms, an issue we aim to address in the present work.

The QKD algorithm is a member of the quantum subspace diagonalization (QSD) family, early examples of which include the quantum subspace expansion of McClean~\textit{et al.}~\cite{mcclean:2017qse} and the imaginary-time quantum Lanczos of Motta~\textit{et. al.}~\cite{motta2020determining}.
In QSD, the eigenpair problem is solved by classically diagonalizing a subspace Hamiltonian constructed in a pre-defined non-orthogonal basis, the matrix elements of which are measured using a quantum device.
This can be done by measuring additional operators to build matrix elements for excitations out of a reference, and/or by explicitly applying a family of unitary generators to form the subspace.
The former category includes techniques such as the aformentioned quantum subspace expansion~\cite{mcclean:2017qse,mcclean2020decoding,takeshita2020increasing,yoshioka2022generalized} and quantum equation-of-motion methods~\cite{ollitrault2020quantum, ganzhorn2019gate, gao2021applications, barison2022quantum}. 
The later category contains a significant amount of variety, but can loosely be partitioned into techniques that employ some form of chemically-inspired unitary ansatz, such as the non-orthogonal variational quantum eigensolver (with~\cite{huggins2020non} or without~\cite{baek2022say} optimization of circuit parameters), and those which employ an ansatz based on (real or imaginary) time evolution. 
As insightfully summarized by Klymko~\textit{et. al.}~\cite{Klymko:2022vp}, at long evolution times, imaginary-time QSD~\cite{motta2020determining,yeter2020practical} can be used to systematically suppress the presence of excited eigenstates, while real-time QSD~\cite{parrish2019quantum,stair:Krylov2020,seki:2021qpm,Klymko:2022vp,Cohn:2021cdf,cortes2022:qk_es,cortes2022fast,shen2022real} removes the presence of excited states by canceling out their phases, similarly to the spirit of classical filter diagonalization~\cite{neuhauser1990bound,neuhauser1994circumventing,wall1995extraction,mandelshtam1997low}. 
In the short-time domain, the time-evolution operator generates a basis that spans a classical Krylov space, highlighting the premise for quantum Krylov diagonalization.  
Techniques for QSD based on eigenvector continuation~\cite{francis2022subspace}, and Davidson diagonalization~\cite{tkachenko2022quantum} have also recently appeared in the literature.

The real-time evolution based QKD algorithm, which we will focus on exclusively in this work, generates a basis through discrete time steps of the time evolution operator, $e^{-i\hat{H}\tau}$, first proposed by Parrish \textit{et al.} as a quantum filter diagonalization algorithm~\cite{parrish2019quantum} and Stair \textit{et al.} as a multi-reference selected quantum Krylov algorithm~\cite{stair:Krylov2020}. 
In recent years, several advances have been made in terms of understanding the theoretical underpinnings of the algorithm \cite{Klymko:2022vp, epperly2022theory, cortes2022:qk_es, shen2022real}, as well as reducing the circuit depth requirements \cite{Cohn:2021cdf}. 
In all cases, the compilation of the real-time evolution operator has relied on a deterministic Trotter-Suzuki decomposition which results in gate depths that are well beyond the reach of current hardware. 
In concrete terms, first order Trotter methods will have a depth scaling $\mathcal{O}(L)$ where $L$ refers to the number of Hamiltonian terms.
Initial gate count estimates based on naive Jordan-Wigner encoding of the second-quantized electronic structure Hamiltonian results in $\mathcal{O}(n_{\mathrm{orb}}^4)$ scaling while low-rank factorized encodings reduce that number to $\mathcal{O}(n_{\mathrm{orb}}^2)$ \cite{Cohn:2021cdf,Motta_2021}, where $n_{\mathrm{orb}}$ is the number of orbitals. 
When considering large-scale molecular systems ($n_{\mathrm{orb}}>50$), these requirements lead to depth estimates on the order of millions for the former and tens of thousands for the latter. 
An outstanding challenge remains in reducing the gate depth requirement in order to make real-time-evolution-based QKD amenable to near-term hardware. 

In this manuscript, we aim to improve this problem by combining the strengths of several orthogonal techniques.
The resultant family of methods, which we refer to as randomized quantum Krylov diagonalization (rQKD) algorithms, provides a systematic way of solving the eigenpair problem on near-term devices by
leveraging the advantages of (i) randomized compilers in the short-time limit, and (ii) reduced scaling afforded by low-rank Hamiltonian factorization.
Our work builds of the rapid progress of stochastic compilation techniques such as Campbell's qDRIFT~\cite{Campbell:2019qdrift} and follow-up work~\cite{kivlichan2019phase, ouyang2020compilation, chem:2021concentraion, wan2022randomized, cho2022doubling}, which have shown how to remove the dependence on the number of Hamiltonian terms from the gate complexity of the time-evolution unitary.
We begin by deriving an error bound for low-rank Hamiltonian factorization used in conjunction with stochastic compilation, and show that one can remove contributions to the error for any target Hamiltonian term.  
We also show that these results can further be improved by performing importance sampling based on analytically-derived optimal weights.
Furthermore, we demonstrate the performance of rQKD numerically via state-vector simulation using a family of hydrogen chains ranging from 6 to 14 atoms and a naphthalene molecule, incorporating finite shot sampling of the measured quantities.
In the present implementation, we find that rQKD is a powerful tool for ground state eigenpair determination, predicting slightly less accurate energies than standard Trotterized QKD with circuits over an order of magnitude shallower.   
Compared to QPE, which requires precise circuit compilation to estimate the phase of the time-evolved unitary $e^{-i\hat{H} \tau}$, the rQKD method leverages the variational principle such that precision requirements are relaxed and noise robustness is improved.
These are crucial features which make rQKD a potential option in the near-term hardware era~\cite{preskill2018quantum}.  

\section{Quantum Krylov method}

Quantum Krylov subspace algorithms aim to solve the eigenvalue equation, $\hat{H}\ket{\psi_k} = E_k\ket{\psi_k}$ where $\ket{\psi_k}$ is the $k$th eigenstate of interest and $E_k$ is the $k$th eigenvalue. 
In the real-time QKD framework (using an evenly spaced time-grid of steps $\Delta \tau$), the variational wavefunction, $\ket{\tilde{\psi}_k}$, is written as a linear combination of non-orthogonal time-evolved quantum states, \begin{equation}
    \ket{\tilde{\psi}_k} = \sum_{n=0}^{D-1} c_n^{(k)} \ket{\phi_n} =  \sum_{n=0}^{D-1} c_n^{(k)} e^{-i\hat{H} n \Delta \tau}\ket{\phi_o}
\end{equation}
where $c_n^{(k)}$ are variational parameters and $\ket{\phi_o}$ is an initial reference state, such as the Hartree-Fock state. The variational coefficients are determined by minimizing the functional,
\begin{equation}
    \mathcal{L} = \braket{\tilde{\psi}_k|\hat{H}|\tilde{\psi}_k} - \lambda( \braket{\tilde{\psi}_k|\tilde{\psi}_k} -1 ),
\end{equation}
with respect to the variational coefficients, $c_n^{(k)}$, yielding the generalized eigenvalue equation,
\begin{equation}
   \mathbf{H}\mathbf{c}^{(k)} = E_k\mathbf{S}\mathbf{c}^{(k)}.
   \label{generalized}
\end{equation}
Here, $\mathbf{c}^{(k)}  = [c_o^{(k)}, c_1^{(k)},\cdots,c_{D-1}^{(k)} ]^T$, represents a column vector of the variational coefficients for the $k$th eigenvalue of interest, while the overlap matrix $\mathbf{S}$ and Hamiltonian subspace matrix $\mathbf{H}$ are defined by the matrix elements,
\begin{align}
    [\mathbf{S}]_{mn} = \braket{\phi_m|\phi_n} \;\;\text{and}\;\;
    [\mathbf{H}]_{mn} = \braket{\phi_m|\hat{H}|\phi_n}.
\end{align}
The hybrid quantum-classical algorithm consists of using the quantum computer to estimate the overlap and Hamiltonian matrix elements using, for instance, Hadamard test quantum circuit measurements and then using the classical computer to solve the generalized eigenvalue problem. The result is an estimate of the eigenvalue $E_k$ and coefficients $c_n^{(k)}$ that provide an approximation to the $k$th eigenstate of interest. 

This procedure can continue iteratively until a stopping criterion is met, however, it also possible to estimate the maximum number of time steps that will be required for convergence by noting that this method builds an order-$D$ Krylov subspace, $\mathcal{K}_D = \text{span}\{\ket{\phi_o},e^{-i\hat{H} \Delta \tau}\ket{\phi_o}, \cdots, e^{-i\hat{H}(D-1) \Delta \tau}\ket{\phi_o} \}$, to define the approximate solution of the eigenvalue problem. 
As shown in \cite{stair:Krylov2020}, the above quantum Krylov space spans the classical Krylov space defined with respect to powers of the Hamiltonian operator in the small-time limit. For a classical Krylov subspace constructed with powers of the Hamiltonian operator, recent work based on the canonical orthogonalization procedure has shown that the Krylov subspace dimension required to predict the ground-state energy $E_0$ with error, $\epsilon_0 \equiv \tilde{E}_0 - E_0 \geq 0$, is given by \cite{epperly2022theory,kirby2022exact}
\begin{equation}
    D \leq \mathcal{O}(\text{min}(\tfrac{1}{\Delta_1},\tfrac{1}{\epsilon_0})\log(\tfrac{1}{\epsilon_0}))
\end{equation}
where $\Delta_1=E_1-E_0$ denotes the spectral gap between the ground and first excited-state eigenvalue of the Hamiltonian. This bound shows that the maximum Krylov dimension $D$ required to achieve an error $\epsilon_0$ can display an inverse dependence on the first excited-state spectral gap and a logarithmic dependence on the inverse of the desired precision when $\Delta_1 > \epsilon_0$. In the limit that an ideal Krylov subspace is implemented on the quantum computer, the maximum gate depth would be much smaller (assuming $\Delta_1 \gg \epsilon_0$) than the depth required for conventional quantum phase estimation algorithm which scales as $\epsilon_0^{-1}$. The trade-off in gate depth reduction, however, comes at the cost sampling complexity $\mathcal{O}(\epsilon^{-2})$ where the Heisenberg scaling is effectively lost. It is worth noting that the inverse dependence on the first-excited state spectral gap is similar to the gap dependence highlighted in recent work \cite{wang2022quantum}, and both are in fact, quadratically worse than the classical Lanczos convergence theory results by Kaniel, Paige and Saad which predict an inverse square root dependence \cite{golub2013matrix}. For quantum Krylov methods, this discrepancy can be explained by the noise-robustness afforded by the canonical orthogonalization procedure which ensures the stability of the generalized eigenvalue problem, Eq. \eqref{generalized}.

In this work, we consider real-time quantum Krylov methods for near-term devices where the implementation of the time-evolution operator acquires a compilation error, and the classical Lanczos bound will not strictly hold. Nevertheless, the convergence bound from above is insightful in explaining a wide variety of numerical experiments which have demonstrated the fast convergence of real-time quantum Krylov methods. 
The purpose of this manuscript will be to study the gate depth incurred by the circuit compilation step, however, before proceeding, it is worth discussing how the total run-time of the hybrid quantum-classical algorithm scales with the Krylov subspace dimension, $D$. 
In this regard, the overlap matrix $\mathbf{S}$ will require $D(D-1)/2$ separate runs to estimate all of the matrix elements, while the Hamiltonian matrix $\mathbf{H}$ will require $L D^2$ separate measurement runs, where $L$ is the number of Hamiltonian terms. 
While the measurements for $[\mathbf{S}]_{mn}$ can be obtained during the evaluation for the $L$ terms of $[\mathbf{H}]_{mn}$, the number of measurements will be dominated by the asymptotic scaling of $\mathcal{O}(LD^2/\epsilon^2)$.

\subsection*{Toeplitz structure}
In the limit that an exact time evolution operator is used, the overlap and Hamiltonian matrix elements take the form,
\begin{equation}
    [\mathbf{S}]_{mn} = \braket{\phi_o|e^{-i\hat{H}(n-m)\Delta\tau}|\phi_o} 
    \label{eq:S_toeplitz}
\end{equation}
\begin{equation}
    [\mathbf{H}]_{mn} = \braket{\phi_o|\hat{H}e^{-i\hat{H}(n-m)\Delta\tau}|\phi_o} 
    \label{eq:H_toeplitz}
\end{equation}
which gives rise to a Toeplitz structure for both matrices. 
In total, the number of measurements required to reconstruct all of the matrix elements will scale as $\mathcal{O}(D)$ and $\mathcal{O}(LD)$ for $\mathbf{S}$ and $\mathbf{H}$ respectively. 
However, the cost of the reduced run-time comes at the price of a high-precision compilation of the time evolution operator, $e^{-i\hat{H}(n-m)\Delta\tau}$. 
As such, in the present manuscript we elect not to implement the near-exact time evolution required for the Toeplitz structure, and instead focus on low-depth and preservation of variationality.
The resulting algorithms then maintain the asymptotic measurement scaling of $\mathcal{O}(LD^2)$.
We point the reader to the appendix of~\cite{parrish2019quantum} which explores the breakdown of variationality when heavily Trotterized circuits are used to evaluate matrix elements of the form given in Eqs.~\eqref{eq:S_toeplitz} and~\eqref{eq:H_toeplitz}.

\begin{figure}[t!]
\centering
\includegraphics[width=3.35in]{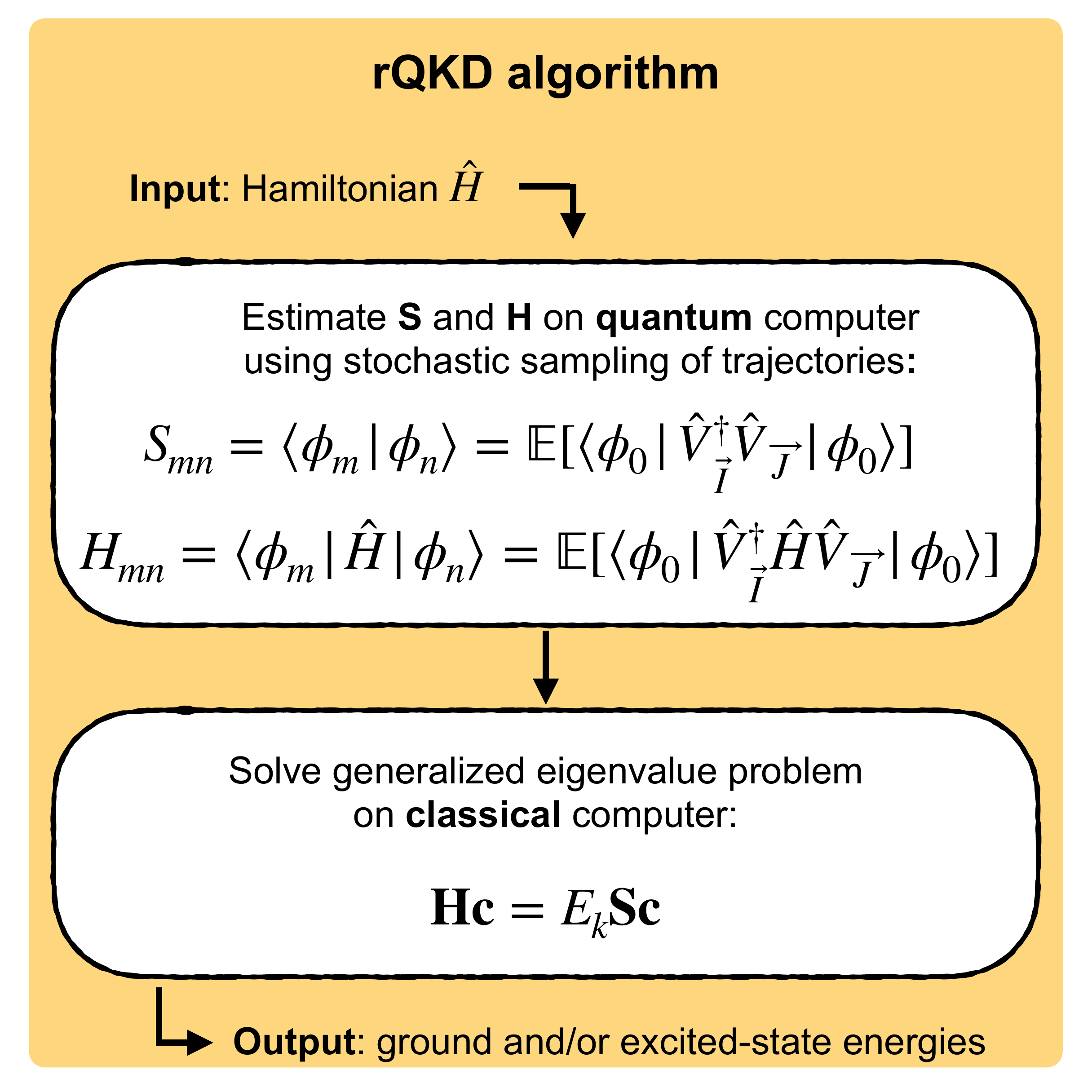}
\caption{Illustration of the rQKD procedure.}
\label{fig:rqkd_scheema}
\end{figure}

\section{Deterministic time evolution}
The main burden of real-time QKD algorithms is based on the circuit compilation step that is required to construct the time-evolution operator, $e^{-i\hat{H}\tau}$. As we show below, the conventional approach for circuit compilation uses deterministic product formulas with resource requirements that are out of reach for near-term hardware. To see why, let us consider a standard Hamiltonian written as a sum of $L$ independently fast-forwardable operators, 
\begin{equation}
    \hat{H} = \sum_s^L \hat{H}_s
    \label{eq:Hamiltonian}
\end{equation}
with terms $\hat{H}_s$ that do not generally commute with one another and have individual spectral-norms defined by, $\lambda_s = \| \hat{H}_s \|$, equal to the maximum singular value. For reasons that will become apparent, we also define $\lambda = \sum_s \lambda_s$, which bounds the spectral norm of the Hamiltonian, $\|\hat{H}\| \leq \lambda$, based on the triangle inequality. Based on the Hamiltonian in Eq.~\eqref{eq:Hamiltonian}, the exact time evolution unitary will be given by, 
$\hat{U}(\tau) = e^{-i \tau \sum_s \hat{H}_s}$. To compile this operator as a quantum circuit, a simple yet effective strategy consists of using the first order Trotter-Suzuki formula 
\begin{align}
    \hat{S}_1(\tau) &= \bigg[ \prod_s^{L} \hat{V}_s (\tau/R) \bigg]^R, 
\end{align}
where it is assumed that each of the unitary sub-components, $\hat{V}_s(\tau) = e^{-i\hat{H}_s\tau}$, can be compiled exactly without additional error. 
The total compilation error, $\epsilon_\mathrm{TS1} \equiv \normx{e^{-i\hat{H}\tau} - \hat{S}_1(\tau)}$, may be truncated to second order in $\tau$ using a Taylor series expansion,
\begin{equation}
    \epsilon_\mathrm{TS1} \approx \frac{\tau^2}{2R} \Big\| \sum_{s' > s}^L \left[\hat{H}_s,\hat{H}_{s'} \right] \Big\| \leq \frac{\lambda^2 \tau^2 }{R}.
\end{equation}
where we have included the spectral norm bound on the right hand side which holds more generally. We note that while tighter commutator bounds exist \cite{childs2021theory}, these bounds can still remain significantly larger than state-dependent compilation errors, $\|\bra{\phi_o} e^{-i\hat{H}\tau}-\hat{S}(\tau) \ket{\phi_o}\|$, which we consider in this manuscript. 
It is also worth noting that for any real state $\ket{\phi_o}$, the state dependent first order Trotter-Suzuki error is zero, and will actually scale to  third order in $\tau$.       

Writing the number of Trotter steps $R$ in terms of $\epsilon_\mathrm{TS1}$, this bound yields a total query complexity, $Q_\mathrm{TS1} = LR$, for the first order Trotter-Suzuki method,
\begin{equation}
    Q_\mathrm{TS1} \approx \frac{L\tau^2 }{2 \epsilon} \Big\| \sum_{s'>s}^L \left[\hat{H}_s,\hat{H}_{s'} \right] \Big\| \leq  \frac{L\lambda^2 \tau^2 }{\epsilon} .
\end{equation}
In other words, this quantifies a bound on the depth of Trotterized quantum circuit. It is important to note that higher ($k$th) order Trotter-Suzuki (TSk) decompositions have been shown to have more favorable query complexity, with upper-bound scaling that approaches linear in $\tau$ and $\lambda$.
The exponential increase in prefactor $\mathcal{O}(5^k)$ (also referred to as the number of stages of the decomposition), however, has caused 2nd and 4th order decompositions to be generally considered most efficient~\cite{childs2021theory}. 

\section{Stochastic time evolution}

A long-standing objective in improved quantum simulation has been the reduction (or removal) of $L$ from the query complexity since $L$ can have up to quartic in the number of orbitals $n_\mathrm{orb}$ for general electronic-structure Hamiltonians.
Inspired by observations that randomization could be beneficial for the $L$ dependence~\cite{childs2019faster}, Campbell developed the quantum stochastic drift protocol~\cite{Campbell:2019qdrift} (qDRIFT) which successfully removed the $L$ dependence altogether at the cost of quadratic $\tau$ scaling and explicit dependence on the norm of the Hamiltonian. The qDRIFT procedure approximates the short time-step evolution operator $\hat{U}(\Delta\tau)$ as the average of a randomly sampled unitary,
\begin{equation}
    e^{-i\hat{H}\Delta\tau}\approx \hat{C}( \Delta\tau) \equiv  \mathbb{E}[\hat{V}( \Delta\tau)] = \sum_s p_s \hat{V}_s(\Delta\tau),
\label{eq:rqk_base}
\end{equation}
which is defined with respect to the discrete probability distribution $\{p_s\}$ satisfying $\sum_s p_s = 1$. Long-time dynamics  can be approximated by multiple powers of this operator, written as:
\begin{equation}
    [\hat{C}(\tau/R)]^R = \mathbb{E}[\hat{V}_1]\cdots \mathbb{E}[\hat{V}_R] = \mathbb{E}[\hat{V}_1\cdots \hat{V}_R],
\label{eq:rqk_rn}
\end{equation}
where the second equality arises from the condition that multiple iterations are independent from one another. For a single instance, this quantity is equivalent to
the product of $R$ randomly sampled unitaries, 
\begin{equation}
    \tilde{U}_\mathrm{qdrift}^{\vec{R}}(\tau) = \prod_r^R \hat{V}_{s_r}(\tau \lambda / R), 
    \label{eq:uqd}
\end{equation}
where the trajectory is indicated by $\vec{R} = \{s_1, s_2, \dots, s_R \}$, a vector of the sampled $s$ indices. We note that the time step $\tau \lambda / R$ is used in order to assure cancellation of the zeroth and first order error in $\tau$ (see the appendix of~\cite{Campbell:2019qdrift} for details). Each $\hat{V}_{s_r}$ term enters into the product above with probability $p_{s_r} = \lambda_{s_r} / \lambda$, biasing the procedure such that $\tilde{U}_\mathrm{qdrift}^{\vec{R}}(\tau)$ tends towards $\hat{U}(\tau)$ as $R$ increases \cite{chem:2021concentraion}. The key feature, after analyzing the difference between the quantum channel for exact evolution and that of qDRIFT, is that the number of terms $R$ (equal to the query complexity) is independent of $L$. The qDRIFT error is explicitly given by:
\begin{equation}
    \epsilon_\mathrm{qdrift} \approx \frac{\tau^2 }{2R} \Big\|  \Big(\sum_{s} \hat{H}_s\Big)^2  -  \sum_s \hat{H}_s^2/p_s \Big\| \leq \frac{2\lambda^2\tau^2 }{R}
\end{equation}
where the second inequality arises from a bound on the diamond norm of quantum channels. As before, we have also included the truncated Taylor series error since we found it still greatly over-estimates the true compilation error in a wide variety of numerical tests.
Rearranging for $R$, the qDRIFT query complexity is then given as:
\begin{equation}
    Q_\mathrm{qdrift} \approx \frac{ \tau^2}{ 2 \epsilon} \Big\|  \Big(\sum_{s} \hat{H}_s\Big)^2  -  \sum_s \hat{H}_s^2/p_s \Big\|  \leq \frac{ 2\lambda^2\tau^2 }{  \epsilon}.
\label{eq:qd_comp}
\end{equation}
Similar to the Trotter-Suzuki result, this bound quantifies the depth of the stochastically sampled time evolution circuit. 
The sample complexity of this procedure required to reproduce the exact time evolution with $\epsilon$ accuracy will scale as $\mathcal{O}(1/\epsilon^2)$ due to the Chernoff bound \cite{Campbell:2019qdrift,wan2022randomized}.

\section{Stochastic Quantum Krylov  Protocol}
With all of the above tools defined, we now formulate a general framework for stochastic real-time quantum Krylov methods. The starting point consists of using the qDRIFT-inspired ansatz, $\hat{C}(\tau)=\sum_{s} p_{s} \hat{V}_{s}(\tau)$,
defined for arbitrary evolution times $\tau$ with respect to the unitary,
\begin{equation}
    \hat{V}_{s}(\tau) = e^{-i \tau \hat{H}_{s}/p_{s}}.
\end{equation}
It is important to note, as will be discussed in future sections, that it is not required that $\hat{V}_{s}$ contain only a single Hamiltonian term, nor a single product. One of the main contributions of this manuscript consists of our proposal of certain combinations of products which lead to reduced error expressions. Furthermore, we consider cases where the sampled unitary consists of a Hamiltonian sub-term with a spectral-norm that is not normalized, $\| H_{s} \|\neq 1$, requiring optimal stochastic weights $p_{s}$ that are fundamentally different from the conventional qDRIFT weights, $\lambda_{s}/\lambda$.

To provide a unified framework for stochastic real-time quantum Krylov methods, we define the order-$D$ quantum subspace,
\begin{equation}
    \mathrm{span}\{\ket{\phi_o},\hat{C}(\tau/R_1)^{R_1}\ket{\phi_o},\cdots, \hat{C}(\tau/R_{D-1})^{R_{_{D-1}}}\ket{\phi_o} \}
\end{equation}
where $\{R_1,R_2,\cdots,R_{D-1}\}$ are integers which correspond to the Krylov basis states and the number of Trotter slices per basis state ($r$) such that $R_n = nr$. We will use this notation in the results section. We should also note that this quantum subspace can span the classical Krylov subspace in the short-time limit \cite{stair:Krylov2020} up to an additive error, therefore a similar convergence behavior will be expected \cite{epperly2022theory}.

As shown in the quantum Krylov section, a general variational ansatz may be written as a linear combination of these states with variational coefficients, $c_n$, which are determined by solving the generalized eigenvalue problem defined by $\mathbf{H}$ and $\mathbf{S}$. These subspace matrix elements can now be written explicitly as the expectation value of the set of random operators, 
\begin{equation}
S_{mn} = \braket{\phi_o|\hat{C}^{\dagger R_m}\hat{C}^{R_n}|\phi_o} = \mathbb{E}[\langle \phi_0|\hat{V}_{\vec{R}_m}^\dagger \hat{V}_{\vec{R}_n}|\phi_0\rangle],
\label{eq:rqkd_smn}
\end{equation}
and,
\begin{equation}
H_{mn} = \braket{\phi_o|\hat{C}^{\dagger R_m}\hat{H}\hat{C}^{R_n}|\phi_o} =  \mathbb{E}[\langle \phi_0|\hat{V}_{\vec{R}_m}^\dagger\!\!\!\hat{H} \hat{V}_{\vec{R}_n}|\phi_0\rangle].
\label{eq:rqkd_hmn}
\end{equation}
We note that in QKD, one must not only consider the query complexity for applying the chosen approximate time evolution circuits, but also the sample complexity required to determine the value of a matrix elements to within sampling error, $\epsilon_\mathrm{m}$ which scales as $\mathcal{O}(1/\epsilon_\mathrm{m}^2)$. 
As we show in the numerical experiments section, it is possible to retain the same $\mathcal{O}(1/\epsilon_\mathrm{m}^2)$ scaling by combining the bitstring measurements of different trajectories \eqref{eq:uqd} for a particular matrix element measurement, $H_{mn}$ or $S_{mn}$. 

We should emphasize that we have written everything in a manner that emphasizes the flexibility afforded by this approach. 
As we shall highlight below, there remains much freedom in the various forms one may chose to define $\hat{V}_s$ which ultimately lead to different runtimes and gate depth requirements. 
We also emphasize again that the aim of the stochastic quantum Krylov approach is to retain variationality and low circuit depth in favor of the exact Toeplitz structure present for exact time evolution. 

\section{Hamiltonian Representation}
Up to this point, the analysis has been given in terms of a general Hamiltonian defined only with respect to $L$ fast-forwardable components. 
It is important to note that while our proposed methodology is applicable to a wide class of Hamiltonians, we consider the number-conserving electronic structure Hamiltonian relevant to chemistry and condensed matter physics,
\begin{equation}
    \hat{H} = E_\mathrm{o} + \sum_{pq,\sigma} h_{pq}\hat{a}^\dagger_{p,\sigma} \hat{a}_{q,\sigma} + \tfrac{1}{2}\!\!\!\sum_{pqrs,\sigma\tau} g_{pqrs} \hat{a}^\dagger_{p\sigma}\hat{a}^\dagger_{r\tau} \hat{a}_{s\tau}\hat{a}_{q\sigma}.
    \label{electronic_hamiltonian}
\end{equation}
Here, $\hat{a}_p^\dagger/\hat{a}_p$ are fermionic creation/annihilation operators of a particle defined with respect to the $p$th spatial orbital; $\sigma,\tau$ are used as labels for the spin of the particle.  
The matrix elements $h_{pq}$ and $g_{pqrs}$ are likewise defined with respect to the spatial orbitals in Appendix A.

For purposes of formulating the rQKD algorithm presented in this work, we will focus on the low-rank, explicit double-factorization (XDF) encoding scheme.
Because it is often used as a sub or superscript, in this manuscript we used XDF (rather than X-DF, used in previous work~\cite{Cohn:2021cdf}) for notational clarity.  
In the XDF formalism, the electronic structure Hamiltonian \eqref{electronic_hamiltonian} is given in the following form
\begin{equation}
\begin{split}
\hat H_\mathrm{XDF} &= \mathcal{E}_0 + \hat{H}_o + \sum_t^{n_\mathrm{DF}} \hat{H}_t \\
&= \mathcal{E}_0 - \frac{1}{2} \sum_{k} f_{k}^\varnothing \hat G^{\dagger}_\varnothing
\left ( \hat Z_{k} + \hat Z_{\bar k} \right ) \hat G_\varnothing \\
 &+ \frac{1}{8} \sum_{t}^{n_\mathrm{DF}} \sum_{k\neq l} Z_{kl}^{t} \hat G^{\dagger}_t
\left ( \hat Z_{k} + \hat Z_{\bar k} \right ) \left ( \hat Z_{l}
+ \hat Z_{\bar l} \right ) \hat G_t,
\label{eq:final_df_ham}
\end{split}
\end{equation}
where $\hat{G}_\varnothing$ and $\hat{G}_t$ correspond to Givens orbital rotation circuits.
The operators $\hat{Z}_{k}$ and $\hat{Z}_{\bar{k}}$ are Pauli Z gates acting on the $k$th and $(k+n_{\mathrm{orb}})$th qubits, respectively. 
The bar denotes the staggered indexing for qubits arranged to correspond to alpha then beta blocks (see Appendix A for details).
The quantity $n_{\mathrm{DF}}$ is the number of Hamiltonian factors with eigenvalues $h_t$ greater than some user-specified threshold $\sigma_\mathrm{DF}$, retained in the first eigen-decomposition of the electron repulsion integral (ERI) tensor, $ (pq|rs) \approx \sum_t^{n_{\mathrm{DF}}} A_{pq}^t h_t A_{rs}^t$. 
The elements $Z_{kl}^t$ are then determined from a subsequent eigen decomposition of $A_{pq}^t = \sum_k U_{pk}^t U_{kq}^t \gamma_k^t$ as $Z_{kl}^{t} = h_t \gamma_k^t \gamma_l^t$, where $\gamma_j^t$ are eigenvalues of $A_{pq}^t$. 
We note that that the XDF Hamiltonian expressed in Eq.~\eqref{eq:final_df_ham} groups effective scalar and one-body contributions from the two-body operator into $\mathcal{E}_0$ and $f_{k}^\varnothing$, respectively.   
A detailed derivation of the XDF Hamiltonian from the conventional second quantized form, including the re-grouping, is given in Appendix A. 
It is also important to note that at larger system sizes one may want to use a Cholesky or density fitting factorization to either directly construct the intermediate XDF tensors, or use them to reconstruct the eigen-decomposition following techniques from classical quantum chemistry~\cite{kallay2014systematic}. 

In the XDF framework, the $\ell_1$ norm, $\lambda_\mathrm{XDF}$, is given by $\lambda_\mathrm{XDF} = \lambda_1 +\lambda_2$, where
\begin{align}
    \lambda_1  &= \sum_{k} |f_k^\varnothing| \\
    \lambda_2 &= \frac{1}{2}\sum_{t kl} |Z^t_{kl}| - \frac{1}{4}\sum_{t k} |Z^t_{kk}|.
\end{align}
We also note that the analogous expressions $\lambda$ using the Jordan-Wigner encoding of the Hamiltonian can be found in~\cite{koridon2021orbital}.

\subsection*{Single-Depth ansatz}
In the simplest case of combining rQKD with double factorization, the unitaries $\hat{V}_s$ can be chosen to correspond to the fast-forwardable operators,
\begin{equation} 
\label{eq:vs_ansatz}
\hat{V}_{s}^{\mathrm{XDF}(1)} (\tau) = \hat{G}_s^\dagger e^{-i\hat{D}_{s} \tau/p_s} \hat{G}_s
\end{equation}
where the superscript $(1)$ indicates an effective-depth of 1 oracle query. As in previous sections, $\hat{D}_s$ refers to a diagonal operator consisting of a summation of either 1-qubit or 2-qubit Pauli Z operators.
For clarification, using the first order XDF unitaries we then build the rQKD ansatz
$\hat{C}^{(1)}_{\mathrm{XDF}}(\tau) = \sum_{s} p_s \hat{V}_{s}^{\mathrm{XDF}(1)}(\tau)$,
noting that the sum over $s$ includes the XDF one-body term $\hat{H}_o$ and two-body terms $\hat{H}_t$.
As a reference point, it is worth calculating the error of this ansatz with respect to the ideal real-time evolution operator. 
For the low-depth XDF(1) ansatz above, this is approximated to second order in $\tau$ as,
\begin{align}
    \epsilon^{\mathrm{XDF}(1)} &= \| e^{-i\hat{H}\tau} - \mathbb{E}[V^{\mathrm{XDF}(1)}(\tau/R)]^R \| \\
    &\approx \frac{\tau^2}{2R} \Big\| \Big(\sum_{s} \hat{H}_s\Big)^2  -  \sum_s \hat{H}_s^2/p_s \Big\|.
\label{eq:v1_error}
\end{align}
The second order approximation is bound by
\begin{equation}
\label{eq:2nd_order_error_bound_d1}
    \epsilon_2^{\mathrm{XDF}(1)} \leq \frac{(\lambda_1 + \lambda_2)^2\tau^2 }{2R}.
\end{equation}

\subsection*{Triple-Depth ansatz}
We also consider the following interleaved form for the unitaries:
\begin{equation} 
\label{eq:vs_ansatz}
\hat{V}_s^{(3)} (\tau) = e^{-i \hat{H}_{s'} \tau / 2} e^{-i \hat{H}_s \tau / p_s} e^{-i \hat{H}_{s'} \tau / 2},
\end{equation}
where $s \neq s'$.
The flexibility of this unitary is significant in how it manifests in the second order error. 
Using $\hat{V}_s^{(3)}$ in Eqs.~\eqref{eq:rqk_base} and~\eqref{eq:rqk_rn}, any choice of weights will produce an ansatz with approximate error $\epsilon^{(3)} = \| e^{-i\hat{H}\tau} - \mathbb{E}[\hat{V}^{(3)}]^R \|$, given to second order in $\tau/R$ as,
\begin{equation}
    \epsilon_2^{(3)} = \frac{\tau^2}{2R} \Big\| \Big(\sum_{s\neq s'} \hat{H}_s\Big)^2  -  \sum_{s \neq s'} \hat{H}_s^2/p_s \Big\|
\label{eq:v3_error}
\end{equation}
where the summation runs over all indices excluding $\hat{H}_{s'}$. 
In other words, Eq.~\eqref{eq:rqk_base} is re-defined in the interleaved unitaries picture so that it excludes the contributions from the $s=s'$ term.
This suggests that the sub-term $\hat{H}_{s'}$ should be chosen as the one with the largest spectral norm or expectation of $\bra{\phi_o} \hat{H}_{s'}^2 \ket{\phi_o}$, in order to reduce the approximation error. 
In the results section, we will discuss several strategies that can be employed to take advantage of this flexibility but will save a thorough investigation of this feature for future studies.
We note that use of Eq.~\eqref{eq:vs_ansatz} results in a similar approach to that used in~\cite{ouyang2020compilation,hagan2022composite} as well as an error expression similar to that recently proposed by~\cite{rajput2022hybridized}, in which Trotterization and qDRIFT are combined using an interaction picture of quantum simulation.
We note that more detailed derivation of error expressions in this section is given in Appendix B.

In the case that the XDF one-body term contains the largest spectral-norm contribution (or the largest value of $\bra{\phi_o} \hat{H}_{s'}^2 \ket{\phi_o}$), then a natural choice for $\hat{H}_{s'}$ would be $\hat{H}_{s'} = - \frac{1}{2} \sum_{k} f_{k}^\varnothing \hat G^{\dagger}_\varnothing \left ( \hat Z_{k} + \hat Z_{\bar k} \right ) \hat G_\varnothing$.
Under this premise and using the XDF Hamiltonian, we propose the following interleaved triple-depth XDF(3) ansatz,
\begin{align}
    \hat{V}_s^{\mathrm{XDF}(3)}(\tau) &= \hat{G}^\dagger_o e^{-i \hat{D}_o \tau / 2} \hat{G}_{os}  e^{-i \hat{D}_s \tau / p_s} \hat{G}_{so} e^{-i \hat{D}_o \tau / 2} \hat{G}_o, 
\end{align}
where $\hat{G}_{os} = \hat{G}_{so}^\dagger = \hat{G}_o \hat{G}_s^\dagger$ may be combined into a single Givens rotation operator. 
Again, now using the XDF-3 unitaries, we construct the rQKD ansatz as 
$\hat{C}_{\mathrm{XDF}}^{(3)}(\tau) = \sum_{s \neq o} p_{s} \hat{V}_s^{\mathrm{XDF}(3)} (\tau)$,
noting that the sum over $s$ does \textit{not} include the XDF one-body term,  $\hat{H}_o$. Additionally, if one chooses probabilities $p_s = \| \hat{H}_s \| / \lambda_2$, then the compilation error will be proportional to square of the spectral norm of the two body operator only, and the second order error is bound by
\begin{equation}
\label{eq:2nd_order_error_bound}
    \epsilon_2^{\mathrm{XDF}(3)} \leq \frac{\lambda_2^2\tau^2 }{2R}.
\end{equation}

\begin{figure*}[ht!]
\centering
\includegraphics[width=7.0in]{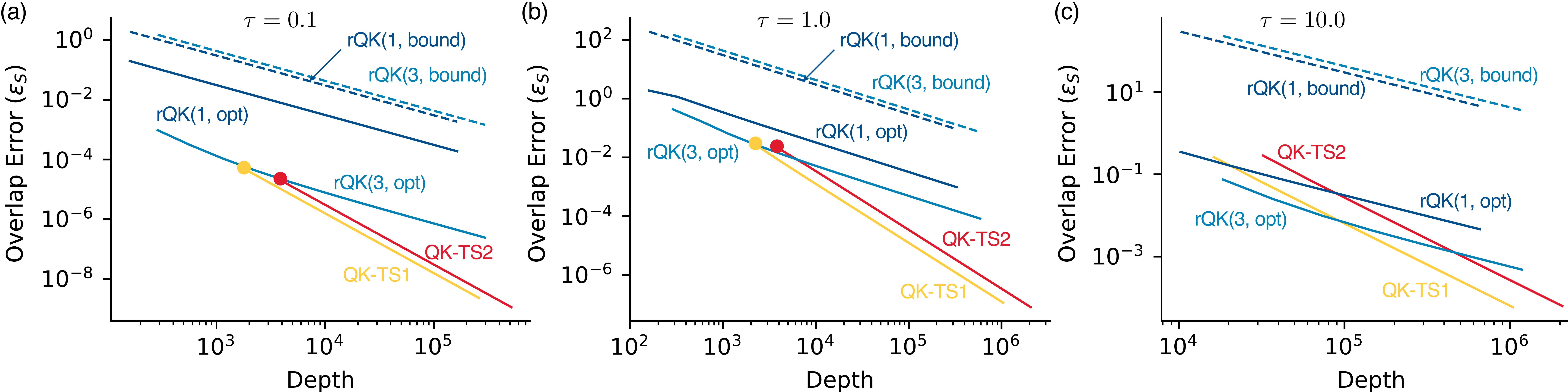}
\caption{Convergence of different overlap matrix element errors, $\epsilon_{S} = | \bra{\phi_0} \hat{U}(\tau) \ket{\phi_0} - \bra{\phi_0} [ \tilde{U}(\delta \tau) ]^{R} \ket{\phi_0}|$, where $\hat{U}(\tau)$ is the exact time evolution operator, $\tilde{U}(\delta \tau) = \hat{C}(\delta \tau), \hat{S}_1(\delta \tau), \hat{S}_2(\delta \tau)$, and $\delta \tau = \tau / R$.
Results are shown for three different time durations [$\tau = 0.1, 1.0, 10.0$ in (a), (b), and (c), respectively] as a function of depth.
The depth is given as a function of increasing Trotter steps $R$ [see Appendix E] for the \ce{H8} Hamiltonian. 
Data sets labeled rQK($k$, opt) indicated rQK using $\hat{C}_{\mathrm{XDF}}^{(k)}$ with $k=1$ or $k=3$ and optimized probability coefficients.
Data sets labeled rQK($k$, bound) indicated the upper bound to the second order error [Eqs.~\eqref{eq:2nd_order_error_bound} and~\eqref{eq:2nd_order_error_bound_d1}] using corresponding numerical values for $\lambda_1$, $\lambda_2$, $\tau$, and $R$.
Caps on the TS curves in (a) and (b) indicate the minimum depth using a single Trotter step $R=1$.}
\label{fig:matrix_element_error_with_queries}
\end{figure*}

Despite the second order bounds in Eqs.~\eqref{eq:2nd_order_error_bound_d1} and~\eqref{eq:2nd_order_error_bound} not accounting for higher-order terms, we provide numerical evidence that they over-estimate the error produced in a quantum Krylov matrix element, often by several orders of magnitude as highlighted in Fig.~\ref{fig:matrix_element_error_with_queries} and discussed in the results section.  

We note that in the conventional Pauli representation where $\hat{H}_s$ represents individual Pauli words multiplied by some coefficient ($\lambda_s \hat{P}_s$), using the above  interleaved unitary would not seem likely to provide a notable advantage. 
This suggests that, in principle, grouped Pauli words that are fast-forwardable should chosen to represent an analog to $\hat{H}_o$ to take full advantage of the interleaved ansatz, however, we leave such studies to future work. 
The explicit double factorization procedure, on the other hand, naturally partitions the Hamiltonian into different groups consisting of the one-body term and the $n_\mathrm{DF}$ separate two-body terms.
As a result, the double factorization procedure naturally takes full advantage of the reduced compilation error afforded by the interleaved ansatz. 

\subsection*{Optimal stochastic weights}
To achieve the lowest second order error [Eq.~\eqref{eq:v1_error} or Eq.~\eqref{eq:v3_error}], we find that it is prudent to optimize the weights so that they minimize the error function with respect to some trial state $\ket{\phi_0}$.
As shown in Appendix C, the analytically optimal probability weights $\tilde{p}_s$ are given by
\begin{equation}
\label{eq:opt_weights}
    \tilde{p}_s = \frac{\sqrt{ \bra{\phi_0} \hat{H}^2_s \ket{\phi_0} }}{\sum_{s'} \sqrt{ \bra{\phi_0} \hat{H}^2_{s'} \ket{\phi_0} }},
\end{equation}
noting that one may wish to exclude terms $\hat{V}_s$ from $\hat{C}$ (and the above expression) for which $\sqrt{ \bra{\phi_0} \hat{H}^2_s \ket{\phi_0} }$ is very small in order to avoid evolution by large $\tau_s$ values.
In practice, we find that this is naturally avoided when the XDF Hamiltonian is truncated to $n_\mathrm{DF} < n^2_\mathrm{orb}$ terms. 

In the situation where $\hat{H}_s$ are simply weighted Pauli-words $(\lambda_s \hat{P}_s)$ such that $\hat{H}_s^2 = \lambda_s^2 \hat{I}$, then the optimal weights become $\tilde{p}_s \rightarrow \lambda_s / \lambda $, equivalent to those used in the original qDRIFT implementation.
However, this is not the case for Hamiltonians with general fast-forwardable subcomponents (such as XDF), and Eq.~\eqref{eq:opt_weights} should be used.
In the following numerical demonstrations section we show in more detail that there is a significant advantage to using optimal weights from Eq.~\eqref{eq:opt_weights} over other options.

We also wish to emphasize that in theory the probabilities could be re-optimized for each time step, however, in practice we find that the initially optimized weights seem seem to be stable across all times from the systems that we studied, and provided little to no advantage to over re-optimized probabilities. Further analysis is needed to fully understand this behavior.

\begin{figure*}[ht!]
\centering
\includegraphics[width=7.0in]{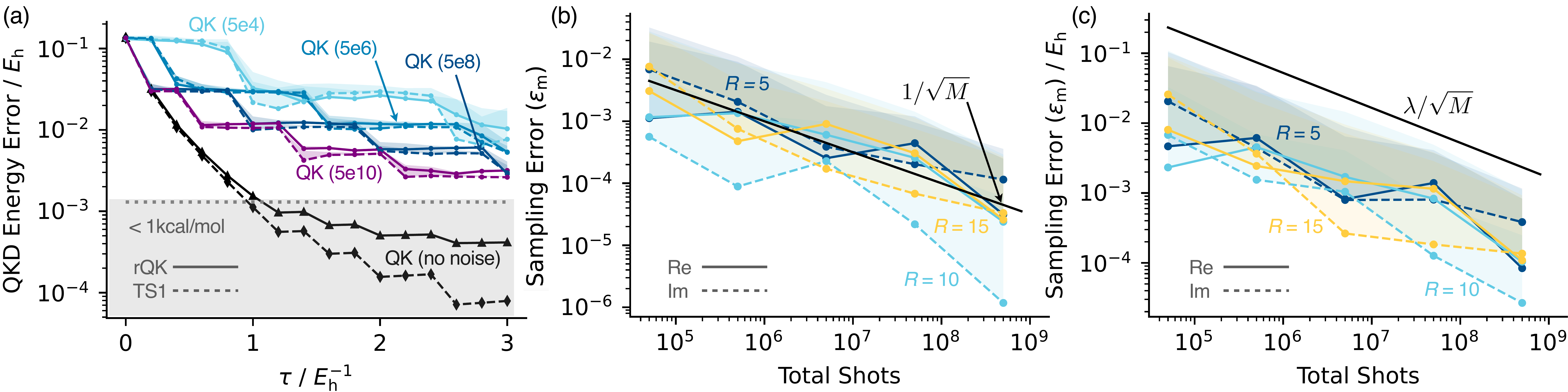}
\caption{Randomized and Trotterized QKD ground state energy convergence with different measurement budgets per matrix element (a), and sampling error $(\epsilon_\mathrm{m})$ in the expectation values for random variables, $\mathbb{E}[ \langle \hat{C}^R (\delta \tau) \rangle ]$~(b) or $\mathbb{E}[ \langle \hat{H} \hat{C}^{R} (\delta \tau) \rangle ]$~(c), as a function of measurement budget.
Energy convergence data sets are reported as rQK($M$), where $M$ is the number of shots used for each matrix element. 
Matrix element errors are relative to the exact expectation value corresponding to an \ce{H8} Hamiltonian with $n_\mathrm{DF}=25$. 
For (a) $\Delta \tau = 0.2$ and $r=5$.
For both (b) and (c), five, ten, and fifteen $(R = nr =5, 10, 15)$ total time steps of length $\delta \tau = 0.04$ were used, corresponding to a full $[\hat{C}_{\mathrm{XDF}}^{(3)}]^R$ operators with $25^5, 25^{10},$~and $25^{15}$ terms, respectively. 
The theoretical sample errors [$1 / \sqrt{M}$ for (a), and $\lambda / \sqrt{M}$ for (b)] are plotted for reference.
All data with noise is averaged over 10 runs with standard deviations indicated by shaded regions above the corresponding data sets.}
\label{fig:shot_noise_single_matrix_element}
\end{figure*}

\begin{figure*}[ht!]
\centering
\includegraphics[width=7.0in]{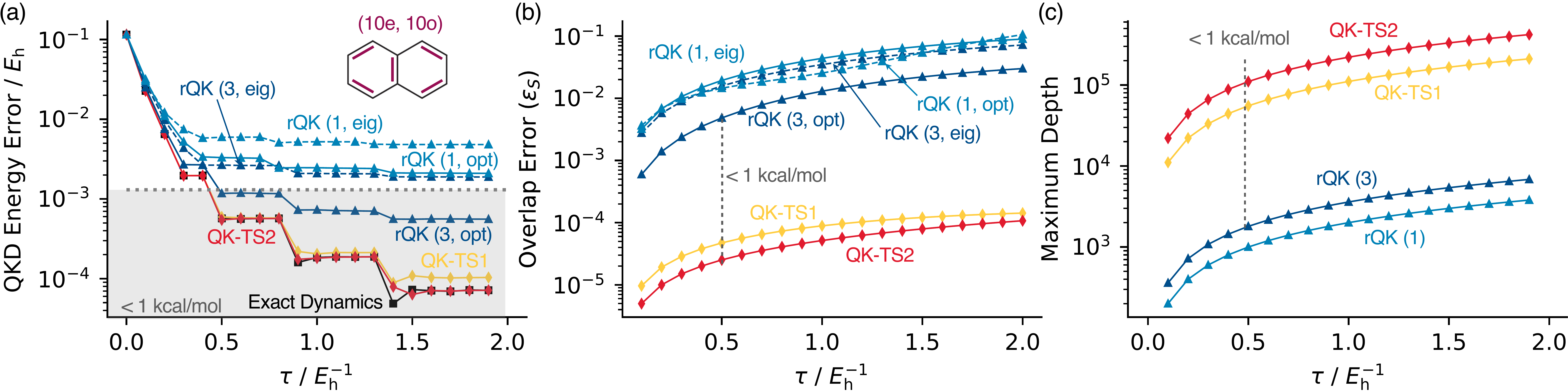}
\caption{Convergence of the QK singlet ground state energy (a), error in the first-row QK overlap matrix elements ($\epsilon_S$) with respect to exact dynamics (b), and the maximum circuit depth required at time $\tau$ (c). All results are for naphthalene in a (10e, 10o) active space, with increasing number of quantum Krylov basis states ($D$).
Here, rQK($k$, $X$) refers to randomized quantum Krylov with $r=2$ Trotter slices, an effective depth of $k=1,3$, and weights determined via either $X=$ first factorization eigenvalue magnitude (eig) [$p_t = h_t / \sum h_t$], or $X=$ optimized weights via Eq.~\eqref{eq:opt_weights} (opt). 
QK-TS1/QK-TS2 refers to first/second order Trotterized QK, respectively, with a single Trotter slice $r=1$.
All calculations used a time step of $\Delta t = 0.1$~a.u. up to $N=20$ total Krylov basis functions.
Note that canonical orthogonalization was employed (using an eigenvalue threshold of $1.0 \time 10^{-12}$) in order to remove linear dependencies. 
}
\label{fig:qk_econv}
\end{figure*}

\subsection*{Towards a practical implementation} 
Based on the strategies and various ansatze introduced above, we now discuss two different implementations that provide various trade-offs in circuit depth and runtime complexity. 

\paragraph{Toeplitz structure implementation} 
If $R_n$ is defined as the power of $\hat{C}(\tau/R_n)$ required to achieve $\epsilon$ accuracy with respect to the ideal time-evolution operator, $e^{-i\hat{H}n \Delta \tau}$, we naturally arrive to the real-time order-$D$ Krylov subspace defined in the quantum Krylov subsection. 
This subspace would approximate the ideal real-time Krylov subspace up to $\epsilon$ accuracy with the query complexity bound, $Q_\mathrm{qdrift} \leq 2\lambda^2\tau^2/\epsilon$. 
In this limit, the Krylov subspace matrices $\mathbf{H}$ and $\mathbf{S}$  would take on a Toeplitz structure which ultimately reduces the total run-time, requiring $\mathcal{O}(D/\epsilon^2)$ measurements. While the reduced run-time is certainly beneficial, this approach will ultimately inherit the strong dependence on $\lambda^2$ and $\tau^2$ inherent to the qDRIFT framework in terms of circuit complexity. 

\paragraph{Near-term implementation}
In comparison, it is also possible to define a randomized Krylov subspace by choosing $\{R_1,R_2,\cdots, R_{D-1} \} = \{1,2 ,\cdots, D-1 \}$, resulting in a Krylov subspace that has maximum gate depth of $D-1$ multiplied by the cost of implementing the unitary $\hat{V}_s(\tau)$. 
While this subspace no longer approximates the ideal real-time Krylov subspace to $\epsilon$ precision, we found numerical evidence that this approach is able to reach a ground-state energy precision of $\sim 10^{-3}$ with gate depths that are orders of magnitude smaller than the equivalent Toeplitz structure approach. Compared to deterministic product formulas, which at best, would have a minimum depth that scales with the number of terms in the Hamiltonian,  $\mathcal{O}(L)$, this approach does not exhibit such scaling thereby resulting in shorter gate depths compared to deterministic Trotterization. Nevertheless, we observed a trade-off in the convergence rate where we found that that this methodology converged more slowly compared to deterministic Trotter methods. In the following section, we provide a more thorough investigation of the stochastic real-time quantum Krylov method for various chemical systems of interest taking into account real sampling noise and compilation errors.

\section*{Numerical Experiments}

We have performed numerical experiments for XDF electronic structure Hamiltonians with simulations corresponding to 12--28 qubits.
Calculations were performed using an in-house GPU-accelerated, spin and number conserving state-vector emulator.
Hydrogen chain calculations all considered an inter-nuclear separation of 1.0~{\AA}, and used a minimal STO-6G basis with restricted Hartree-Fock orbitals.
Naphthalene calculations used a cc-pVTZ basis with RHF orbitals performed in a (10e, 10o) active space using all $\pi/\pi*$ orbitals. 
The active space was identified using the automatic valence active space procedure~\cite{Sayfutyarova_2017} implemented in the \textsc{PySCF} package~\cite{sun2020recent}.
The naphthalene active space orbitals are also depicted graphically in Appendix D.
Quantum Krylov energy errors for hydrogen chains and naphthalene are reported relative to the full configuration interaction (FCI), or complete active space CI (CASCI) values, respectively.  

Whenever comparing Trotterized QKD and rQKD, XDF Hamiltonians with the same $n_\mathrm{DF}$ are used in both cases with an XDF eigenvalue threshold of $\sigma_\mathrm{DF} = 1.0 \times 10^{-8}$~$E_\mathrm{h}$.
Results in all figures/tables in this section are reported in terms of micro ($\delta \tau$) and macro $(\Delta \tau = r \delta \tau)$ time steps, where $r$ is the number of Trotter-steps per Krylov basis state (hence forth referred to as Trotter-slices for clarity).
The total number of trotter steps for the $n$th Krylov basis state is then $R_n = nr$.

To begin, in Fig.~\ref{fig:matrix_element_error_with_queries} we numerically compare stochastic compilation of the single-depth and triple-depth ansatze against first (TS1) and second (TS2) order Trotter-Suzuki decompositions by examining the absolute error in quantum Krylov overlap matrix element $\epsilon_{S} = | \bra{\phi_0} \hat{U}(\tau) \ket{\phi_0} - \bra{\phi_0} [ \hat{C}(\tau/R) ]^{R} \ket{\phi_0}|$ and $\epsilon_{S} = | \bra{\phi_0} \hat{U}(\tau) \ket{\phi_0} - \bra{\phi_0} [ \hat{S}_k(\tau/R) ]^{R} \ket{\phi_0}|$, respectively as a function of circuit depth (see Appendix E for details on the constant-factor gate complexity estimates) for TS1, TS2, and the randomized quantum Krylov protocol.  
Fig.~\ref{fig:matrix_element_error_with_queries} (a-c) shows $\epsilon_{S}$ for the \ce{H8} Hamiltonian at three time durations: $\tau = 0.1, 1.0, 10.0$~$E_\mathrm{h}^{-1}$, representative of errors relative to exact dynamics.

In Fig.~\ref{fig:shot_noise_single_matrix_element} we demonstrate the robustness of sampling the random variable expectations using many individual measurements of states generated using the rQK ansatz. 
Fig.~\ref{fig:shot_noise_single_matrix_element} (a) plots the rQK and TS1-QK ground-state energy convergence as a function of evolution time at different shot budgets, specifying the number of shots ($M$) used for each matrix element.
Importantly, in Fig.~\ref{fig:shot_noise_single_matrix_element} (a), canonical orthogonalization eigenvalue thresholds ($\sigma_\mathrm{CO}$) one order or magnitude larger than the expected measurement error were used in order to dampen the effects of the shot noise, (i.e. $\sigma_\mathrm{CO} = 10 \times M^{-1/2}$).
A value of $\sigma_\mathrm{CO} = 10^{-12}$ was used for rQK without shot noise. 
Fig.~\ref{fig:shot_noise_single_matrix_element} (b) and (c) show the convergence of the measured values for Eq.~\eqref{eq:rqkd_smn} and~\eqref{eq:rqkd_hmn} (corresponding to $[\hat{C}_{\mathrm{XDF}}^{(3)}]^R$ using an \ce{H8} Hamiltonian) with an increasing number of shots, and different vales of $R$. 
Within the the present framework, each shot is specified by a bitstring corresponding to a single determinant $\ket{\Phi_I}$ as well as an ancilla value.
The state from which the bitstring is drawn is a normalized linear combination of the bra and ket states with the corresponding givens rotation $\hat{G}_s$ applied such that the contribution of that shot to the overall expectation value is given by $\bra{\Phi_I} \hat{D}_s \ket{\Phi_I}$. 
A more detailed overview of this procedure can be found in Appendix G.

In Fig.~\ref{fig:qk_econv} we show the naphthalene ground state energy convergence, first-row overlap matrix error (relative to exact dynamics), and maximum circuit depth with respect to the total amount of time evolution for TS and randomized quantum Krylov.
For QK-TS1 and QK-TS2 we plot results using the minimum $r=1$ Trotter slices per Krylov basis state with a time step of $\Delta \tau = 0.1$~$E_\mathrm{h}^{-1}$.
For randomized QK we use $r=2$ Trotter slices per Krylov basis state with a time step of $\Delta \tau = 0.1$~$E_\mathrm{h}^{-1}$ with both the single-depth ($\hat{C}_{\mathrm{XDF}}^{(1)}$) and triple-depth interleaved ansatze ($\hat{C}_{\mathrm{XDF}}^{(3)}$), each with optimized [Eq.~\eqref{eq:opt_weights}] and eigenvalue [$p_t = h_t / \sum h_t$] weights). 
We note that the plateaus in convergence are the result of systematically removing linear dependencies in the Krylov basis via canonical orthogonalization.
Details of this procedure can be found in Appendix F.

Finally, in Table \ref{tab:resource_estimates} we compare QKD energy errors and computational resource estimates for chains of 6 to 14 hydrogens arranged on a line with an inter-nuclear separation of 1.0~{\AA}.
These systems encompass $n_\mathrm{DF}$ values (determined using a threshold $\sigma_\mathrm{DF}=10^{-8}$~$E_\mathrm{h}$) ranging from 18 (for \ce{H_6}) to 48 (for \ce{H_14}), and Hilbert-space sizes ranging from $4.9 \times 10^3$ to $1.2 \times 10^7$.  
Here we focus on the circuit depth and ground-state energy errors using TS1, TS2, and rQK using both the $\hat{C}_{\mathrm{XDF}}^{(1)}$ and $\hat{C}_{\mathrm{XDF}}^{(3)}$ ansatze.

\begin{table*}[!ht]
\centering
\caption{Computational resources and singlet ground state energy errors ($\Delta E_\mathrm{X}$ / m$E_\mathrm{h}$) using rQK and 1st-order Trotterized QK (QK-TS1) for linear chains of 6-14 hydrogen atoms.
The value $n_{\mathrm{H}}$ refers to the number of hydrogen atoms and $N_{\rm{FCI}}$ is the full dimension of the corresponding Hilbert-space. 
The quantity $d_\mathrm{X}$ indicates the maximum circuit depth required at any point in the calculation and X=(rQK or QK-TS1). 
All calculations used a time step of $\Delta \tau = 0.1$~$E_\mathrm{h}^{-1}$, two Trotter slices ($r=2$), and seven Krylov basis states ($D = 7$), with the exception of \ce{H6} which used only six to improve numerical stability. 
Results here do not use canonical organization in order to achieve the fastest convergence possible for comparison purposes. 
} 
\footnotesize 

\begin{tabular*}{\textwidth }{@{\extracolsep{\stretch{1.0}}}*{3}{c}*{6}{r}@{}}
    \hline
    \hline
     $n_\mathrm{H}$  & $ n_\mathrm{DF} $  & $N_{\rm{FCI}} $  & $d_\mathrm{rQK(d3)}$ & $\Delta E_{\rm{ rQK(d3) }}$ & $ d_{\rm{ rQK(d1) }}$ &   $ \Delta E_{\rm{ rQK(d1) }} $ & $d_\mathrm{QK-TS1}$ & $\Delta E_{\rm{ QK-TS1 }}$ \\
    \hline
    
  6  & 18  &  $4.9 \times 10^3$   &  1296 &  0.505 & 720  &  1.325 &   13248  & 0.347  \\
  8  & 25  &  $6.3 \times 10^4$   &  2016 &  1.055 & 1120 &  4.505 &   28448  & 0.630  \\
  10 & 33  &  $4.0 \times 10^2$   &  2520 &  2.705 & 1400 & 11.883 &   46760  & 1.128  \\
  12 & 41  &  $8.5 \times 10^5$   &  3024 &  5.196 & 1680 & 25.109 &   69552  & 2.107  \\
  14 & 48  &  $1.2 \times 10^7$   &  3528 &  8.849 & 1960 & 48.192 &   94864  & 3.177  \\[6pt]
 
    \hline
    \hline
\end{tabular*}
\label{tab:resource_estimates}
\end{table*}

\section*{Discussion}

In this section we discuss the numerical results presented above.
To begin, in Fig.~\ref{fig:matrix_element_error_with_queries} (c), at the longest duration [$ \tau = 10.0$~$E_\mathrm{h}^{-1}$], there is a clear crossover point at $\epsilon_S \approx 10^{-2}$ below which the TS1 decomposition performs more favorably than stochastic compilation.
At $\tau = 0.1,1.0$, we see that the crossover points are near the minimum ($R=1$) depth for TS1 and TS2, near errors of approximately $\epsilon_S \approx 10^{-4} $ at $\tau = 0.1$~$E_\mathrm{h}^{-1}$, and $\epsilon_S \approx 10^{-2}$ at $\tau = 1.0$~$E_\mathrm{h}^{-1}$.
We generally observe the expected trend that for a given error the relative advantage of stochastic compilation diminishes with increasing $\tau$.
It is also worth pointing out that at short and medium evolution times, $\tau = 0.1$ and $\tau=1.0$, rQK offers a way to generate relatively accurate matrix elements with a depth significantly less than the minimal depth ($R=1$) first order and second order Trotter-Suzuki deterministic compilation strategies. 
This is the key advantage of the stochastic compilation technique, and as discussed below, results in the ability of rQK to produce accurate ground state energy predictions (errors less than 1~kcal~mol$^{-1}$) with sub TS1 gate depth. 
Additionally, we find that the numerical value for $\epsilon_{S}$ is, at all three durations, orders of magnitude lower than the second order upper bound given in Eqs.~\eqref{eq:2nd_order_error_bound_d1} and~\eqref{eq:2nd_order_error_bound}. 
The discrepancy can likely be attributed to numerous cancellations of terms which will naturally occur when evaluating the many Pauli Z expectations present in Eqs.~\eqref{eq:v1_error} and~\eqref{eq:v3_error}. 

In analyzing the results shown in Fig.~\ref{fig:shot_noise_single_matrix_element} (a) we see that, using $\Delta \tau = 0.2$ and $r=5$, rQK and QK-TS1 (in the absence of shot noise) rapidly converge to errors below 1~kcal~mol$^{-1}$.
As mentioned in the previous section, however, we also find that as the number of measurements is decreased it is necessary to increase the amount of information discarded during canonical orthogonalization in order to maintain variationality. 
As a result, the energy convergence for rQK and QK-TS1 with shot noise is significantly slower than in the noise-free scenario. Interestingly, there is little to no distinction in the performance of rQK and QK-TS1 up to the limit of $5 \times 10^{10}$ shots per matrix element and noise-abatement canonical orthogonalization protocol. Additionally, we observe that the convergence for rQK matrix elements shown in in Fig.~\ref{fig:shot_noise_single_matrix_element} (a) and (b) is in good agreement with the statistical estimates of $M_S = \mathcal{O}(1/\epsilon_\mathrm{m}^2$), and $M_H = \mathcal{O}(\lambda/\epsilon_\mathrm{m}^2$). 
We also observe that there is no significant change in convergence or standard deviation when increasing the number of Trotter steps, noting that $R=5,10,15$ correspond to $\hat{C}^{(3)}_\mathrm{XDF}$ operators with $25^5$, $25^{10}$, and $25^{15}$ total trajectories. 

Based on the results presented in Fig.~\ref{fig:qk_econv}, we find that rQKD is able to converge almost as quickly as 1st and 2nd-order Trotterized QKD until a total evolution time of $\tau = 0.4$~$E_\mathrm{h}^{-1}$ (five quantum Krylov basis states at $\Delta \tau = 0.1$~$E_\mathrm{h}^{-1}$).
Notably rQKD using $\hat{C}_{\mathrm{XDF}}^{(3)}$ and optimal weights [rQK(3, opt)], QK-TS1, QK-TS2, and exact dynamics all converge below 1~kcal~mol$^{-1}$ (1.29~m$E_\mathrm{h}$) with a total evolution time of only $\tau = 0.5~E_\mathrm{h}^{-1}$.
This is significant because the maximum circuit depth required by rQK(3, opt) is orders of magnitude smaller than those required by QK-TS1 and QK-TS2. 
Moreover, sub 1~kcal~mol$^{-1}$ energy accuracy can be achieved with rQK at a depth less than a single TS1 Trotter step. 
Other variants of rQKD converge relatively slowly after $\tau = 0.5~E_\mathrm{h}^{-1}$, but there is still a significant improvement seen for both $\hat{C}_{\mathrm{XDF}}^{(1)}$ and $\hat{C}_{\mathrm{XDF}}^{(3)}$ when using the optimal probability weights from Eq~\eqref{eq:opt_weights}.

Perhaps the most significant feature of Fig.~\ref{fig:qk_econv} can be seen in comparing QK-TS and rQK in (a) vs (b). 
As noted above, both techniques achieve sub 1~kcal~mol$^{-1}$ energy accuracy with respect to FCI, but the matrix element error for rQK is numerically 1-2 orders of magnitude larger than the energy error with respect to exact dynamics. 
This behavior indicates that, while reproducing exact dynamics accelerates convergence in QK methods, it is not essential to the prediction of highly accurate eigenvalues. 
While this merits further analysis left for future work, we believe that success in the inexact-dynamics domain is afforded by the classical Krylov-like basis generated by the rQK ansatz in the short-time limit.   

For the hydrogen chains in Tab.~\ref{tab:resource_estimates}, similarly to the the case of naphthalene shown in Fig.~\ref{fig:qk_econv}, we find that in general rQKD does not converge as quickly as first order Trotterized QKD, particularly when the $\hat{C}^{(1)}_\mathrm{XDF}$ ansatz is used. 
Interestingly, we observe that for all flavors of QKD, the errors in the ground state energy have (approximately) quadratic scaling with respect to the number of Krylov basis states ($D$) when using a fixed number of Krylov basis states and Trotter slices.
Importantly, we also observe a dramatic reduction in both total circuit depth and circuit depth scaling relative to Trotterized QKD.
For example, with the same values of $D$ and $r$, \ce{H14} can be treated using rQKD circuit with maximum depth equal to $3.53 \times 10^3$, while Trotterized QKD requires a maximum depth of $9.49 \times 10^5$.  

\section*{Summary and Outlook}

In this manuscript, we have proposed a new class of stochastic quantum Krylov diagonalization algorithms aimed at solving the eigenstate estimation problem in quantum many-body physics and quantum chemistry. 
By combining stochastic compilation techniques, low-rank Hamiltonian factorization, and the variationality inherent to quantum Krylov methods, we obtained a highly robust algorithm with $\mathcal{O}(n_\mathrm{orb})$ scaling per Trotter step, providing a quadratic improvement in depth compared to deterministic approaches.

Within the framework of double factorization, we derive a bound for the factor $\lambda$ which in turn bounds the second order error expressions for stochastic compilation, and present an ansatz which for which target Hamiltonian terms can be omitted from the error bounds. 
We also have shown numerically that our ansatz produces errors several orders of magnitude smaller than the derived second order upper bound. 

We demonstrated the performance of our rQKD approach with realistic state-vector simulations of various molecular systems including linear hydrogen chains ranging from 6 to 14 atoms, as well as napthalene in an active spaces of (10e, 10o), showing that convergence below 1~kcal~mol$^{-1}$ errors can be achieved with a very small number of time steps and circuit depths achievable in the near-term hardware era (orders of magnitude shallower than deterministic low-rank Trotterization).
We also found in numerical comparisons that stochastic compilation enables simulation in the regime where a single deterministic Trotter step is too expensive, and is even favorable in general above certain error thresholds or in the short-time regime.
In this context, our preliminary estimates suggest that rQKD algorithm could be immediately deployed with current hardware for simulating systems on the order of (14e,14o),
making it a viable candidate for using quantum simulation to treat systems non-trivial sizes.  

The improvements highlighted in this manuscript can be attributed to the fact that deterministic high-order Trotter methods, while achieving depth linear in time $\tau$, also scale linearly with the number of Hamiltonian terms ($L$). 
Randomized compilation techniques, on the other hand, can (in many circumstances) achieve similar accuracy with depths that are independent of $L$, but quadratic in $\tau$. 
Ultimately, the $L$-independence as well as the quadratic $\tau$ dependence makes randomized compiling ideally suited for quantum Krylov methods which inherently aim to solve the eigenpair problem in the short-time limit.

Like many other variational quantum algorithms~\cite{peruzzo2014variational,yung2014transistor,mcclean2016theory}, QKD approaches exploit an inherent trade-off between circuit depth and sampling complexity. 
This will likely lead to much longer run-times compared to deterministic approaches based on quantum phase estimation, highlighting that there is no free lunch when it comes to quantum algorithmic design.
Our results also indicate that if long time evolution or very high accuracy are desired, it is likely more efficient to use deterministic Trotterization or post Trotter methods, similarly to Campbell's original findings. 
As such, we view the present combination of rQK executed with low-rank Hamiltonians as one point on a road-map of feasibly realizable QKD algorithms, likely at a position of higher cost/accuracy that Pauli qDRIFT, but lower cost/accuracy than full deterministic Trotterization with XDF Hamiltonians. 

Like virtually all QSD algorithms, rQKD by default produces eigenpair estimates for $D$ states. 
However, using the present implementation we only explore its application to ground state energy determination.
Based on the Krylov convergence analysis, application of rQKD to excited state will likely be more challenging due to the tendency for the spectral gap $\Delta_k =E_{k+1} - E_k$ to vanish with increasing $k$.  
Previous work on treating excited states with quantum Krylov methods~\cite{parrish2019quantum, cortes2022:qk_es} has also indicated a strong dependence on the choice of initial state $\ket{\phi_o}$, and time step size $\Delta \tau$ used when constructing the space.
As such, we leave thorough investigation of rQKD excited state determination to future work. 

We note also that the techniques used in rQKD for sampling random variables may seem reminiscent of the classical shadows framework introduced by Huang \textit{et. al}~\cite{huang2020predicting}.
However, in the present work, each matrix element constitutes the measurement of a single observable with respect to many different states, rather than a collection of observables measured from a single state. 
Although the classical shadows framework may not be immediately applicable for rQK, both approaches may benefit from a more rigorous combined study in the future.

While this manuscript has highlighted rQKD as a viable route towards solving the eigenpair problem on near-term quantum hardware, future work will investigate possible ways of improving the convergence of the rQKD algorithm using warm-starts, multi-determinantal initialization, and potentially derivations of higher-order representations. 
Further effort should also be directed towards an implementation on real hardware where connectivity and error mitigation techniques will be of key importance. 

\section{ACKNOWLEDGMENTS}
The authors thank William Kirby for sharing his
expertise on the convergence properties of Krylov methods. 
N.S. and C.C. would also like to thank Ed Hohenstein for helpful comments. 
The QC Ware effort in this work was supported by the U.S. Department of Energy, Office of Science, Basic Energy Sciences, Chemical Sciences, Geosciences and Biosciences Division.

N.S., C.C., and R.M.P. own stock/options in QC Ware Corp.

\clearpage

\section*{Appendix A: Double Factorized Hamiltonian}
In spin-orbital notation, the conventional second quantized Hamiltonian in electronic structure theory is given by:
\begin{equation}
    \hat{H} = E_\mathrm{nuc} + \sum_{pq,\sigma} h_{pq}a^\dagger_{p,\sigma} a_{q,\sigma} + \tfrac{1}{2}\sum_{pqrs,\sigma\tau} g_{pqrs} a^\dagger_{p\sigma}a^\dagger_{r\tau} a_{s\tau}a_{q\sigma}
\end{equation}
where $p,q,r,s$ denote spatial orbitals and $\sigma,\tau$ denote the spins of the electrons. Here, $E_\mathrm{nuc}$ denotes the nuclear-nuclear repulsion energy, while the one-electron and two-electron tensors are defined as,
\begin{align}
    h_{pq} &\equiv (p|h|q) \\
    & = \int\! d\mathbf{r}\; \phi_p^*(\mathbf{r})\left(-\frac{1}{2}\nabla^2 - \sum_I \frac{Z_I}{ r_I}\right)\phi_q(\mathbf{r}), \\
    g_{pqrs} &\equiv (pq|rs) \\
    & = \iint d\mathbf{r}_1 d\mathbf{r}_2\; \phi_p^*(\mathbf{r}_1)\phi_q(\mathbf{r}_1)r_{12}^{-1}\phi_r^*(\mathbf{r}_2)\phi_s(\mathbf{r}_2).
\end{align}
The one-electron term contains contributions from one-electron kinetic and nuclear-electron attractive potential while the second term describes the electron-electron repulsion. While this Hamiltonian exactly represents the electronic structure problem, it is often beneficial to work with the spin-free version of the Hamiltonian because it provides a simplification of the expressions. Using the spin-free notation, the electronic structure Hamiltonian is re-written as:
\begin{equation}
\hat H = E_{\mathrm{nuc}} +
\sum_{pq} h_{pq} \hat E_{pq} +
\frac{1}{2}\sum_{pqrs}g_{pqrs} ( \hat E_{pq} \hat  E_{rs} - \delta_{qr} \hat{E}_{ps} )
\end{equation}
where the spin-summed singlet one-particle substitution operator is, $\hat E_{pq} \equiv \hat{a}_{p}^{\dagger}\hat{a}_{q}
+ \hat{a}_{\bar p}^{\dagger} \hat{a}_{\bar q}$, where a bar over the spatial orbital indices $p,q$ indicates a $\beta$ spin orbital, and the absence of a bar indicates an $\alpha$ spin orbital. The summation runs over all of the spatial orbitals of the particular problem. In the active space picture, the spatial orbitals are separated into core, active, and virtual orbital contributions. By tracing out the core and virtual space orbitals, the active space Hamiltonian is obtained where only the active space orbitals remain,
\begin{equation}
\hat H_\mathrm{active} \equiv E_{ext} +
\sum_{pq} \kappa_{pq} \hat E_{pq} +
\frac{1}{2}\sum_{pqrs}g_{pqrs} \hat E_{pq} \hat  E_{rs} 
\end{equation}
with normalized coefficients,
\begin{align}
    E_{\mathrm{ext}} &=
E_{\mathrm{nuc}} + 2\sum_{i}^{\mathrm{core}}h_{ii} + \sum_{ij}^{\mathrm{core}}[2g_{iijj} - g_{ijij}] \\
    \kappa_{pq} &= h_{pq} +  \sum_{i}^{\mathrm{core}}[2g_{pqii}
- g_{piqi}] - \frac{1}{2}\sum_{r}g_{prrq}.
\end{align}
From here, it is possible to use various fermion-to-qubit mappings such as the Jordan-Wigner of Bravyi-Kitaev transformations in order to obtain a Hamiltonian that is amenable to qubit-based quantum computing. If we only apply such mappings, then we will obtain a Hamiltonian where the total number of terms will scale as $N^4$ where $N$ corresponds to the total number of qubits. The low-rank, double factorized formulation provides a way of avoiding this scaling by performing a two-step factorization procedure. The first step groups the $pq$ and $rs$ indices of the two-electron integral tensor resulting in the eigendecomposition, 
\begin{equation}
    g_{(pq)(rs)} = \sum_t A_{pq}^t h_t A_{rs}^t.
\end{equation}
We then perform a second factorization of each eigenvector,
\begin{equation}
    A_{pq}^t = \sum_k U_{pk}^t U_{kq}^t \gamma_k^t.
\end{equation}
Substituting these expressions into the active space Hamiltonian, we obtain
\begin{equation}
\begin{split}
    \hat{H} = &E_{ext} + \sum_{pq} \kappa_{pq} \hat E_{pq}  \\
    &+ \frac{1}{2}\sum_{pqrs}\sum_t \sum_{kl} U_{pk}^t U_{kq}^t U_{rl}^t U_{ls}^t (\gamma^t_k h_t \gamma^t_{l}) \hat{E}_{pq}\hat{E}_{rs}. 
\end{split}
\end{equation}
The summation over $p,q,r,s$ orbitals can be interpreted as a transformation of the spatial orbitals which effectively defines new creation and annihilation operators, $\tilde{a}^\dagger_{k t}  = \hat{G}_t^\dagger \hat{a}_k^\dagger \hat{G}_t = \sum_p U_{pk}^t \hat{a}^\dagger_p$, where 
\begin{equation}
\hat{G}_t = \exp\left( \sum_{pq} [\log \mathbf{U}^t]_{pq} \hat{a}^\dagger_p \hat{a}_q \right).
\end{equation}
The resulting  expression for the active space Hamiltonian is:
\begin{equation}
\label{eq:givens_ham1}
\hat H_\mathrm{active} = E_{\mathrm{ext}} + \sum_{pq} \kappa_{pq} \hat E_{pq} + \frac{1}{2} \sum_{t} \sum_{kl} Z_{kl}^{t} \hat G^{\dagger}_t \hat E_{kk}
\hat E_{ll} \hat G_t,
\end{equation}
where we defined $Z_{kl}^{t} = \gamma^t_k h_t \gamma^t_l$.
Before factoring the one-body term, we will perform a fermion-to-qubit mapping using the Jordan-Wigner transformation. Using $\hat{E}_{kk} = I - \tfrac{1}{2}(\hat{Z}_k + \hat{Z}_{\bar{k}})$, we find the following identity,
\begin{equation}
\hat E_{kk}\hat E_{ll} = -\hat I+\hat E_{kk}
+ \hat E_{ll} + \frac{1}{4} \left ( \hat Z_{k}
+ \hat Z_{\bar k} \right ) \left ( \hat Z_{l}
+ \hat Z_{\bar l} \right ),
\end{equation}
where $\hat{Z}_k$ denotes a Pauli $Z$ operator which acts on qubit $k$. This identity shows that it is possible to partition the Hamiltonian once more. The two-body operator is now written in terms of an effective scalar and one-body portion,
\begin{equation}
\begin{split}
& \frac{1}{2}\sum_{t} \sum_{kl} Z_{kl}^{t} \hat G^{\dagger}_t
\hat E_{kk}\hat E_{ll}\hat G_t = \\
&  -\frac{1}{2} \sum_{t} \sum_{kl} Z_{kl}^{t} + \sum_{t} \sum_{k}
\left [ \sum_{l} Z_{kl}^{t} \right ] \hat G^{\dagger}_t
\hat E_{kk} \hat G_t \\
& + \frac{1}{8} \sum_{t} \sum_{kl} 
Z_{kl}^{t} \hat G^{\dagger}_t \left ( \hat Z_{k} +
\hat Z_{\bar k} \right ) \left ( \hat Z_{l} + \hat Z_{\bar l} \right ) \hat G_t.
\end{split}
\end{equation}
Folding the effective terms into the scalar and one-body components through back transformations,  we obtain:
\begin{equation}
\begin{split}
\hat H_\mathrm{active} = & E_{\mathrm{ext}} - \frac{1}{2} \sum_{pq} g_{ppqq} \\
&+ \sum_{k} f_{k}^{\varnothing} \hat{G}^\dagger_{\varnothing}  \hat{E}_{kk} \hat{G}_{\varnothing} \\
&+ \frac{1}{8} \sum_{t} \sum_{kl} Z_{kl}^{t} \hat G^{\dagger}_t \left ( \hat Z_{k} + \hat Z_{\bar k} \right )
\left ( \hat Z_{l} + \hat Z_{\bar l} \right ) \hat G_t.
\end{split}
\end{equation}
where we performed an eigendecomposition of the one-body tensor, $f_{pq} = \kappa_{pq} + \sum_r g_{pqrr}$, such that, $f_{pq} = \sum_k f_k^\varnothing U_{pk}^\varnothing U_{kq}^\varnothing$. After one final re-arrangement of the scalar term, and re-writing the Hamiltonian with respect to Pauli $\hat{Z}$ operators only, we obtain the final form of the double factorized Hamiltonian,
\begin{equation}
\begin{split}
\hat H_\mathrm{active} = &\mathcal{E}_0 - \frac{1}{2} \sum_{k} f_{k}^\varnothing \hat G^{\dagger}_\varnothing
\left ( \hat Z_{k} + \hat Z_{\bar k} \right ) \hat G_\varnothing \\
 &+ \frac{1}{8} \sum_{t} \sum_{k\neq l} Z_{kl}^{t} \hat G^{\dagger}_t
\left ( \hat Z_{k} + \hat Z_{\bar k} \right ) \left ( \hat Z_{l}
+ \hat Z_{\bar l} \right ) \hat G_t,
\label{eq:final_df_ham}
\end{split}
\end{equation}
where the final scalar is defined as $\mathcal{E}_0 \equiv E_\mathrm{ext} + \sum_{k}f_k^\varnothing - \frac{1}{2} \sum_{pq} g_{ppqq} + \tfrac{1}{4}\sum_{tk}Z_{kk}^t$.
In all of the following discussions on implementation, the Hamiltonian defined in Eq.~\eqref{eq:final_df_ham} is what will be referenced.

\section*{Appendix B: Error expressions for randomized time-evolution}

In the following, we derive the time evolution error bounds for the randomized time evolution circuits. 
We first consider the first order randomized expression,
$\hat{C}(\tau) = \sum_s p_s e^{-i\tau\hat{H}_s/p_s}$. The corresponding time evolution error is given by,
\begin{equation}
    \epsilon = \| e^{-i\tau\hat{H}} - [\hat{C}(\tau/R)]^R \|.
\end{equation}
Expanding $\hat{U}(\tau) = e^{-i\tau\hat{H}}$ and $\hat{C}(\tau)$ to second order we obtain 
\begin{equation}
\hat{U}(\tau) \approx 1 - i\tau \hat{H} - \tfrac{\tau^2}{2} \hat{H}^2,    
\end{equation}
and
\begin{equation}
\hat{C}(\tau) \approx 1 - i\tau \hat{H} - \tfrac{\tau^2}{2}\sum_s \hat{H}^2_s/p_s,  
\end{equation}
respectively. 
Using that
\begin{equation}
[\hat{C}(\tau)]^R \approx 1 - iR\tau\hat{H} - R\tfrac{\tau^2}{2}\sum_s \hat{H}_s^2/p_s - \tfrac{R(R-1)}{2}\tau^2 \hat{H}^2,    
\end{equation}
the second order time evolution error (which will dominate in the small $\tau/R$ limit) can be written as,
\begin{equation}
    \epsilon_2 = \frac{\tau^2}{2R} \Big\| \sum_s (p_s^{-1} - 1) \hat{H}_s^2 - \sum_{s\neq s'} \hat{H}_s\hat{H}_{s'}  \Big\|,
\end{equation}
the expression reported in the main text. 
We can bound this expression in terms of the norms $\lambda$ and $\lambda_s$ as
\begin{equation}
    \epsilon_2 \leq \frac{\tau^2}{2R} \Bigg( \sum_s \frac{ \| \hat{H}_s \|^2 }{p_s} + \sum_{s,s'} \| \hat{H}_s \| \| \hat{H}_{s'} \| \Bigg).
\end{equation}
Using qDRIFT probabilities, $p_s = \| H_s \| / \lambda$, we obtain 
\begin{equation}
    \epsilon_2 \leq \frac{\tau^2}{2R} \Bigg( \lambda \sum_s \lambda_s + \sum_{s,s'} \lambda_s \lambda_{s'} \Bigg)
    = \frac{\tau^2 \lambda^2}{R}.
\end{equation}

\subsection*{Error bounds for the triple-depth ansatz}

In the more specific case that the interleaved $\hat{C}_{\mathrm{XDF}}^{(3)}$ ansatz is used, Taylor expanding each term of the LCU gives
\begin{equation}
\begin{split}
    \hat{C}_{\mathrm{XDF}}^{(3)}(\tau) = & \sum_t p_t e^{-i \hat{H}_o \tau / 2} e^{-i \hat{H}_t \tau / p_t} e^{-i \hat{H}_o \tau / 2} \\
    \approx & \sum_t p_t \times \\
    & \Big( 1 - \tfrac{i \tau}{2} \hat{H}_o - \frac{\tau^2}{8} \hat{H}^2_o \Big) \times \\
    & \Big( 1 - \tfrac{i \tau}{p_t} \hat{H}_t - \frac{\tau^2}{2p_t^2} \hat{H}^2_t \Big) \times \\
    & \Big( 1 - \tfrac{i \tau}{2} \hat{H}_o - \frac{\tau^2}{8} \hat{H}^2_o \Big).
\end{split}
\end{equation}
Collecting the zero and one-body terms above gives
\begin{equation}
    \sum_t p_t \Big[ 1 - i \tau (\hat{H}_o + \hat{H}_t ) \Big]
    = 1 - i \tau (\hat{H}_o + \sum_t \hat{H}_t),
\end{equation}
which is equivalent to the exact time evolution up to first order.
Similarly, collecting the second order terms gives the corresponding expression for the error operator
\begin{equation}
\begin{split}
    \hat{E}^{(3)}_\mathrm{XDF} = & \Big[ (\hat{H}_o + \sum_t \hat{H}_t)^2 \\
    - & \sum_t p_t \Big( \hat{H}_o^2 + \frac{\hat{H}_t^2}{p_t^2} + \frac{\hat{H}_t \hat{H}_o}{p_t} + \frac{\hat{H}_o \hat{H}_t}{p_t}  \Big) \Big] \\
    = & \sum_t (p_t^{-1} - 1) \hat{H}^2_t - \sum_{t,t' \neq t} \hat{H}_t \hat{H}_{t'},
\end{split}
\end{equation}
the expression reported in the main text.

\section*{Appendix C: Optimal weighting coefficients}
Based on the linear combination of unitaries ansatz from the main text [Eq.~\eqref{eq:rqk_base}], the optimal weights can be derived by defining the Lagrangian,
\begin{equation}
    \mathcal{L} = \sum_s c_s(1-p_s^{-1})- \lambda( \sum_n p_s -1 ),
\end{equation}
where $c_s = \braket{\phi_0|\hat{H}^2_s|\phi_0}$.
To ensure that the coefficients $p_s$ remain positive, we use the constraint, $p_s = g_s^2$, resulting in, 
\begin{equation}
    \mathcal{L} = \sum_s c_s(1-g_s^{-2}) - \lambda( \sum_s g_s^2 -1 ).
\end{equation}
Taking the partial derivative with respect to $g_s$ and the Lagrange multiplier $\lambda$, we obtain
\begin{align}
    \frac{\partial \mathcal{L}}{\partial g_s} &= 2\frac{c_s}{g_s^3} - 2\lambda g_s, \\
    \frac{\partial \mathcal{L}}{\partial \lambda} &= -\left(\sum_s g_s^2 -1 \right).
\end{align}
At the stationary point, we obtain the relation $c_s = \lambda g_s^4$. Similarly, we have the condition $\sum_s g_s^2 = 1$, which leads to optimal Lagrange multiplier, $\lambda = (\sum_s \sqrt{c_s})^2$. The optimal weighting coefficients are then given by, $g_s^2 = \sqrt{c_s}/\sum_{s}\sqrt{c_s}$.

\section*{Appendix D: Active Space Orbitals for Naphthalene}

\begin{figure}[ht!]
\centering
\includegraphics[width=3.4in]{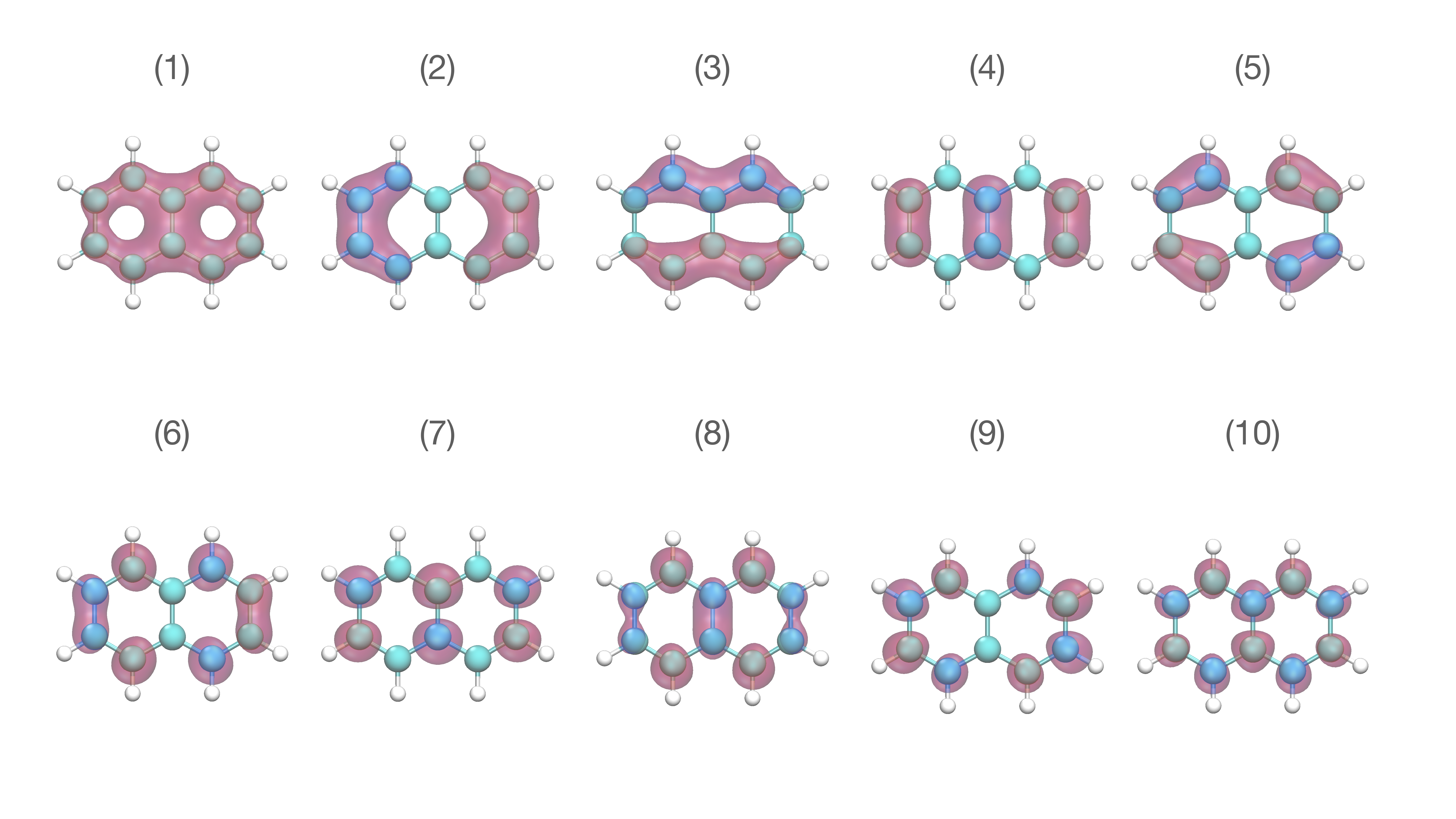}
\caption{Active space orbitals used for naphthalene in this study at a restricted Hartree-Fock cc-PVTZ level of theory. The ten active orbitals show here correspond to the $\pi/\pi*$ space and were selected using the automated valence active space procedure implemented in \textsc{PySCF} to identify $2p_z$ orbitals.}
\label{fig:naphthalene_active_space_orbtilas}
\end{figure}

\section*{Appendix E: Quantum Circuit Construction}
In the following, we provide an explicit quantum circuit reconstruction of the time evolution operator required for the double-factorized active space Hamiltonian, Eq. (\ref{eq:final_df_ham}). First we consider the decomposition for a single deterministic Trotter step where the time evolution operator is approximated by the first order product formula,
\begin{equation}
    \hat{U}(\tau) = e^{-i\hat{H}_\mathrm{active}\tau} = e^{-i\mathcal{E}_0\tau}\hat{G}_o^\dagger \hat{V}_{1} \hat{G}_o\prod_t \hat{G}_t^\dagger \hat{V}_{2}^t \hat{G}_t  + \mathcal{O}(\tau^2)
\end{equation}
where the one-body and two-body diagonal time-evolution operators are given by,
\begin{align}
    \hat{V}_{1} &= e^{i\tau/2 \sum_k f_k^\varnothing(\hat{Z}_k +\hat{Z}_{\bar{k}}) } \label{1body}\\
    \hat{V}_{2}^t &=  e^{-i\tau/8 \sum_{kl} Z^t_{kl} (\hat{Z}_k +\hat{Z}_{\bar{k}})(\hat{Z}_l + \hat{Z}_{\bar{l}}) }. \label{2body}
\end{align}
It is clear from this expression that the only tools needed to provide an explicit quantum circuit implementation of the time-evolution operator are Givens gates required to simulate the orbital rotation operators, $\hat{G}_t$, as well as a circuit implementation of the one-body and two-body diagonal operators, $\hat{V}_{1}$ and $\hat{V}_{2}^t$. We will first outline the decomposition of the orbital rotation operators $\hat{G}$ followed by the decomposition of the one-body and two-body time-evolution operators.

\subsection*{Givens rotations}
Orbital rotations can be efficiently implemented on a quantum computer with linear circuit depth and nearest-neighbor hardware connectivity \cite{Kivlichan:2021ld}. This is achieved by decomposing the total orbital rotation operator as a product of two-body Givens rotations $G_{\phi}$,  $\hat{G}_t = \prod_n \hat{G}_{\phi^t_n}$, where the action of the two-body Givens rotations is to perform an effective QR decomposition of the orbital rotation matrix \cite{wecker2015solving}. A single 2-orbital Givens fabric is represented in matrix form as,
\begin{equation}
    G_\phi = 
    \begin{pmatrix}
1 & 0 & 0 & 0 \\
0 & \cos\phi & -\sin\phi & 0 \\
0 & \sin\phi & \cos\phi & 0 \\
0 & 0 & 0 & 1
\end{pmatrix}
\end{equation}
with a circuit implementation given by:
\[ \Qcircuit @C=1em @R=1em {
& \multigate{1}{G_\phi} & \qw \\
& \ghost{G_\phi} & \qw \\
} 
\hspace{0.2cm}
\Qcircuit @C=1.em @R=.3em {
\\\\\\\\
& \mbox{$\rightarrow$}
}
\hspace{0.5cm}
\Qcircuit @C=1em @R=.7em {
& \qw & \targ & \gate{R_y(\phi)} &\targ  &\qw &\qw  \\
& \gate{H} & \ctrl{-1} & \gate{R_y(\phi)} & \ctrl{-1} &\gate{H} &\qw
} 
\hspace{1.8cm}
\]

\subsection*{One-body evolution}
The one-body evolution is trivially executed with constant $\mathcal{O}(1)$ depth circuits using single-qubit Pauli Z rotation gates, $R_z(\theta) = \exp\left(i\theta\hat{\sigma}_z/2\right)$. The alpha and beta qubits labelled by $k$ and $\bar{k}$ respectively will have the same rotation angle given by, $\theta = -\tau f_k^\varnothing$, for a single time step, $\tau$.

\subsection*{Two-body evolution}
\begin{figure}[ht]
\[ \Qcircuit @C=1em @R=1.em {
& \multigate{3}{e^{i\hat{V}^t_{2}\tau}} & \qw \\
& \ghost{e^{i\hat{V}^t_{2}\tau}} & \qw \\
& \ghost{e^{i\hat{V}^t_{2}\tau}} & \qw \\
& \ghost{e^{i\hat{V}^t_{2}\tau}} & \qw 
} 
\hspace{0.5cm}
\Qcircuit @C=1.em @R=.3em {
\\\\\\\\ \\ \\ \\\\\\\\
& \mbox{$\rightarrow$}
}
\hspace{0.5cm}
\Qcircuit @C=1em @R=1em {
& \multigate{1}{\mathcal{F}_{\theta_{01}}}  &\qw & \multigate{1}{\mathcal{F}_{\theta_{13}}} & \qw & \qw \\
& \ghost{\mathcal{F}_{\theta_{01}}}  & \multigate{1}{\mathcal{F}_{\theta_{03}}} & \ghost{\mathcal{F}_{\theta_{01}}} & \multigate{1}{\mathcal{F}_{\theta_{12}}} & \qw \\ 
& \multigate{1}{\mathcal{F}_{\theta_{23}}}  & \ghost{\mathcal{F}_{\theta_{03}}} & \multigate{1}{\mathcal{F}_{\theta_{02}}} &  \ghost{\mathcal{F}_{\theta_{12}}} & \qw \\
& \ghost{\mathcal{F}_{\theta_{23}}} & \qw &\ghost{\mathcal{F}_{\theta_{02}}} & \qw & \qw \\
} 
\]
\caption{Four-qubit SWAP network.}
\label{fig:swap_circ_4qb}
\end{figure}
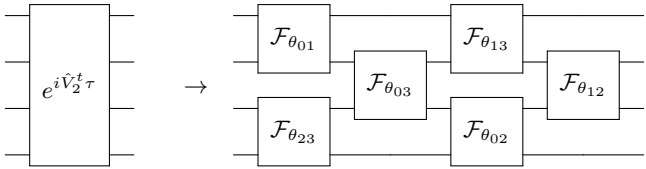

The two-body Z-Z rotation has the following circuit implementation:
\[ \Qcircuit @C=1em @R=.7em {
& \gate{ZZ_\theta}\qwx[1] & \qw \\
& \gate{ZZ_\theta} & \qw
}
\hspace{0.5cm}
\Qcircuit @C=1.em @R=.3em {
\\\\\\\\
& \mbox{$\rightarrow$}
}
\hspace{0.5cm}
\Qcircuit @C=1em @R=1.2em {
& \ctrl{1} & \qw & \ctrl{1} & \qw \\
& \targ & \gate{R_z(\theta)} & \targ & \qw 
} 
\]
Assuming that the hardware has all-to-all connectivity, it is possible to show that exact evolution of the two-body diagonal operator requires $\binom{N}{2}$ of these gate fabrics with a total gate depth of $N-1$. In practice, however, the hardware connectivity is more limited. Assuming linear connectivity, it is preferable to use swap networks which use the following gate fabric
\[ \Qcircuit @C=1em @R=1em {
& \multigate{1}{\mathcal{F}_\theta} & \qw \\
& \ghost{\mathcal{F}_\theta} & \qw \\
} 
\hspace{0.5cm}
\Qcircuit @C=1.em @R=.3em {
\\\\\\\\
& \mbox{$\rightarrow$}
}
\hspace{0.5cm}
\Qcircuit @C=1em @R=1.2em {
& \ctrl{1} & \targ & \qw & \ctrl{1} & \qw \\
& \targ & \ctrl{-1} & \gate{R_z(\theta)} & \targ & \qw 
} 
\]
which can be written in matrix form as,
\begin{align}
    \mathcal{F}_\theta &= 
    \begin{pmatrix} e^{i\theta/2} & 0 & 0 & 0 \\
    0 & 0 & e^{-i\theta/2} & 0 \\
    0 & e^{-i\theta/2} & 0 & 0 \\
    0 & 0 & 0 & e^{i\theta/2}
    \end{pmatrix}.
\end{align}
A SWAP network can then be used to efficiently generate all-to-all interaction. This is achieved by applying this fabric to neighboring qubit pairs in an alternating even and odd pattern. In Fig.~\ref{fig:swap_circ_4qb}, we demonstrate this idea for a system of four qubits.
It is important to note that the qubit order will be reversed after a single application of the SWAP network. Using this approach, it is possible to simulate diagonal two-body evolution with all-to-all interaction using $\binom{N}{2}$ gates and a depth of exactly $N$ with respect to these gate fabrics. 
Combining these ideas, we provide the quantum circuit implementation of a single rank-1 Trotter step (including one-body and two-body terms) for a 4 spin-orbital system in Fig.~\ref{fig:trotter_step_circ_4qb},
\label{fig:trotter_step_circ_4qb}
\begin{figure*}[ht]
\[
\Qcircuit @C=1em @R=1em {
& \multigate{1}{G_{\phi_{01}}} & \gate{Z_{\theta_0}} & \multigate{1}{G_{\phi'_{01}}} & \multigate{1}{\mathcal{F}_{\theta_{01}}}  &\qw & \multigate{1}{\mathcal{F}_{\theta_{13}}} & \qw & \multigate{1}{G_{\phi''_{01}}} & \qw \\
& \ghost{G_{\phi_{01}}} & \gate{Z_{\theta_1}} & \ghost{G_{\phi_{01}}} & \ghost{\mathcal{F}_{\theta_{01}}}  & \multigate{1}{\mathcal{F}_{\theta_{03}}} & \ghost{\mathcal{F}_{\theta_{01}}} & \multigate{1}{\mathcal{F}_{\theta_{12}}} & \ghost{G_{\phi_{01}}} & \qw \\ 
& \multigate{1}{G_{\phi_{01}}} & \gate{Z_{\theta_0}} & \multigate{1}{G_{\phi'_{01}}} & \multigate{1}{\mathcal{F}_{\theta_{23}}}  & \ghost{\mathcal{F}_{\theta_{03}}} & \multigate{1}{\mathcal{F}_{\theta_{02}}} &  \ghost{\mathcal{F}_{\theta_{12}}} & \multigate{1}{G_{\phi''_{01}}} & \qw \\
& \ghost{G_{\phi_{01}}} & \gate{Z_{\theta_1}} & \ghost{G_{\phi_{01}}} & \ghost{\mathcal{F}_{\theta_{23}}} & \qw &\ghost{\mathcal{F}_{\theta_{02}}} & \qw & \ghost{G_{\phi''_{01}}} & \qw \\
} 
\]
\caption{Quantum circuit for a single rank-1 Trotter step for a 4 spin-orbital system.}
\label{fig:trotter_step_circ_4qb}
\end{figure*}
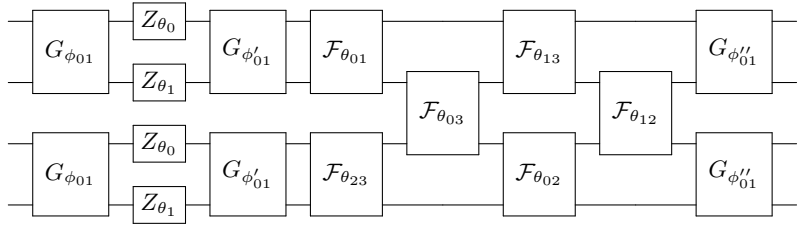
where we used alpha-then-beta ordering where the first two qubits correspond to the $\alpha$ orbitals while the last two qubits correspond to the $\beta$ orbitals.

\subsection*{Gate Complexity Estimate}
The total CNOT count for a single Givens fabric acting on $N$ qubits will be $2\binom{N}{2} = N(N-1)$. On the other hand, the gate count for the all-to-all ZZ interaction will be $3\binom{N}{2} = 3N(N-1)/2$. Using alpha-then-beta ordering for the qubits results in gate fabrics with a total number of $2N(N/2-1) + 3N(N-1)/2 = N/2(5N-7)$ CNOTS for the evolution of a single rank two-body operator. The corresponding CNOT gate depth will be $5N$. The total CNOT gate depth required for the quantum Krylov method will be $5 N R (D-1)$ where $R$ denotes the number of Trotter iterations and $D$ denotes the dimension of the Krylov subspace. 
It is important to note that this gate complexity estimate corresponds to the uncontrolled time-evolution unitary. The controlled unitary operator required for the Hadamard test will require controlled versions of the one-body and two-body evolution operators as discussed in \cite{Cohn:2021cdf} resulting in a constant multiplicative factor increase in the depth.

\section*{Appendix F: Canonical orthogonalization}

Although it routinely employed in quantum chemistry, we feel it is helpful here to give a brief overview of how we use canonical orthogonalization to ameliorate linear dependencies and shot noise in the QK generalized eigenvalue problem.
We first perform an eigendecpmosition of the QK overlap matrix as $\mathbf{S} = \mathbf{U}^\dagger \mathbf{S} \mathbf{U}$, such that $s_{mm}$ is the $i$th eigenvalue of $\mathbf{S}$. 
In our procedure, we use only the truncated set of eigenvalues $ \{ \tilde{s}_{mm} \}$ greater than some user specified threshold $\sigma_\mathrm{CO}$ to form an orthogonalizer $\tilde{\mathbf{X}}$ with matrix elements given by $\tilde{X}_{mn} = U_{mn} / \sqrt{\tilde{s}_{mm}} $.
We then transform the QK Hamiltonian ($\mathbf{H}$) using the orthogonalizer to form $\tilde{\mathbf{H}} = \tilde{\mathbf{X}}^\dagger \mathbf{H} \tilde{\mathbf{X}}$.
Finally, we diagonalize $\tilde{\mathbf{H}}$ to get estimates for the QK energies $E_k$ and (after back-transforming with $\mathbf{X}$) the eigenvector coefficients $c_m^{(k)}$.

The procedure inevitably results in removing some information contained in the origional quantum Krylov basis with the tradeoff of stabilizing the generalized eigenvalue solve.
Numerically, we find that it is imperative to use both when the overlap matrix becomes singular, or when there is shot-nose present in the system. 
We note that for calcluations which include a finite shot budget $M$ we always set $\sigma_\mathrm{CO}$ an order of magnitude above the measurement precision $\epsilon_\mathrm{m} \approx 1 / M^{-1/2}$ to ensure stable results. 

\section*{Appendix G: Determination of matrix elements from individual shots}

Here we will give an overview of how we classically implement the matrix element sampling with individual shots (used to generate the data in Fig.~\ref{fig:shot_noise_single_matrix_element}).
Consider two that we wish to determine both the real and imaginary components of two quantities: (i) $\langle \phi_L | \phi_R \rangle$ and (ii) $\bra{\phi_L} \hat{H}_t \ket{\phi_R}$, where $\hat{H}_t = \hat{G}_t^\dagger \hat{D}_t \hat{G}_t$ (as in the main text) is a single term of the XDF Hamiltonian. 
We consider a single shot to mean an $n_\mathrm{qb} = 2n_\mathrm{orb} + 1$ qubit register readout corresponding to a bitstring which defines both a determinant $\ket{\Phi_I}$ and a single ancilla $\ket{q_I}$.
For a set of $M$ shots, we (simultaneously) calculate the real quantities
\begin{equation}
    \Re [ s ] = \frac{1}{M} \sum_I^{M} \bra{q_I} \hat{Z} \ket{q_I} 
\end{equation}
and,
\begin{equation}
    \Re [v_t] = \frac{1}{n_\mathrm{Re}^0} \sum_I^{n_\mathrm{Re}^0} \bra{\Phi_I} \hat{D}_t \ket{\Phi_I} + \frac{1}{n_\mathrm{Re}^1} \sum_J^{n_\mathrm{Re}^1} \bra{\Phi_J} \hat{D}_t \ket{\Phi_J}, 
\end{equation}
as well as their imaginary counterparts for which $\ket{\Phi_I}$ and $\ket{q_I}$ are drawn from slightly different distributions as described below. 
The quantities $n_\mathrm{Re}^0, n_\mathrm{Re}^1$ [as well as $n_\mathrm{Im}^A, n_\mathrm{Im}^B$] are the number of times the ancilla was measured to be in the $\ket{0}$ vs $\ket{1}$ state, we likewise outline how these quantities are determined below.

To emulate the results from a Hadamard test, we first construct four states given by different linear combinations of $\ket{\phi_L}$ and $\ket{\phi_R}$:
\begin{equation}
    \ket{\phi_x^A} = \ket{\phi_L} + \ket{\phi_R}, 
\end{equation}
\begin{equation}
    \ket{\phi_x^B} = \ket{\phi_L} - \ket{\phi_R}, 
\end{equation}
\begin{equation}
    \ket{\phi_y^A} = \ket{\phi_L} - i\ket{\phi_R}, 
\end{equation}
and
\begin{equation}
    \ket{\phi_y^B} = \ket{\phi_L} + i\ket{\phi_R}, 
\end{equation}
with respective square moduli $N_x^A, N_x^B, N_y^A$, and $N_y^B$.
We then define the states 
\begin{equation}
    \ket{\phi_X} = \ket{\phi_x^A} + \ket{\phi_x^B}, 
\end{equation}
and,
\begin{equation}
    \ket{\phi_Y} = \ket{\phi_y^A} + \ket{\phi_y^B}, 
\end{equation}
similarly with square moduli $N_X$, and $N_Y$.
The readouts for $\ket{q_I}$ are then sampled from a distribution where the probability of measuring $q_I = 0$ for the ancilla qubit is simply $p_\mathrm{Re}(0) = N_x^A / N_X$ for the real contributions and $p_\mathrm{Im}(0) = N_y^A / N_Y$ for the imaginary contributions.
Similarly, the probabilities of measuring $q_I = 1$ are given by $p_\mathrm{Re}(1) = N_x^B / N_X$ for the real contributions and $p_\mathrm{Im}(1) = N_y^B / N_Y$.
With each sample, we accumulate the actual number of times 1 or 0 was measured for the real and imaginary cases as $n_\mathrm{Re}^0, n_\mathrm{Re}^1, n_\mathrm{Im}^0$, and $n_\mathrm{Im}^1$.

We then follow a similar procedure when sampling the main-register readout to determine the $\Phi_I$ bitstrings. 
To do so, we form the (normalized) states $\ket{\tilde{\phi}_X^t} = N_X^{-1/2} \hat{G}_t \ket{\phi_X}$ and $\ket{\tilde{\phi}_Y^t} = N_Y^{-1/2} \hat{G}_t \ket{\phi_Y}$, such that the corresponding wave-function amplitudes [$| C_I^X |^2$, and $| C_I^Y |^2$] form distributions in which the probability of sampling a determinant (bitstring) $\ket{\Phi_I}$ is equal to the corresponding amplitude. 
It is worth noting that (after sorting the vector of probability amplitudes) determinants can be sampled extremely efficiently on a GPU using random number generation and parallel binary searches. 
As such, drawing up to approximately $M = 10^{10}$ samples in routine calculations is manageable for the molecular systems we study in this manuscript. 

We also note that in the main text we make the assumption that matrix elements for normalized quantum Krylov basis states of the form $ \ket{\Psi_R} = [\hat{C}]^R \ket{\phi_0}$ (where $\hat{C} = \sum_s p_s \hat{V}_s$) can be sampled following the procedure outlined above.
This can only be done where because a weighted mixture of Bernoulli distributions is itself a Bernoulli distributions with the same mean \cite{McLachlan2000Finite}.  

\newpage

\bibliography{bib1,bib2}

\begin{thebibliography}{64}%
\makeatletter
\providecommand \@ifxundefined [1]{%
 \@ifx{#1\undefined}
}%
\providecommand \@ifnum [1]{%
 \ifnum #1\expandafter \@firstoftwo
 \else \expandafter \@secondoftwo
 \fi
}%
\providecommand \@ifx [1]{%
 \ifx #1\expandafter \@firstoftwo
 \else \expandafter \@secondoftwo
 \fi
}%
\providecommand \natexlab [1]{#1}%
\providecommand \enquote  [1]{``#1''}%
\providecommand \bibnamefont  [1]{#1}%
\providecommand \bibfnamefont [1]{#1}%
\providecommand \citenamefont [1]{#1}%
\providecommand \href@noop [0]{\@secondoftwo}%
\providecommand \href [0]{\begingroup \@sanitize@url \@href}%
\providecommand \@href[1]{\@@startlink{#1}\@@href}%
\providecommand \@@href[1]{\endgroup#1\@@endlink}%
\providecommand \@sanitize@url [0]{\catcode `\\12\catcode `\$12\catcode
  `\&12\catcode `\#12\catcode `\^12\catcode `\_12\catcode `\%12\relax}%
\providecommand \@@startlink[1]{}%
\providecommand \@@endlink[0]{}%
\providecommand \url  [0]{\begingroup\@sanitize@url \@url }%
\providecommand \@url [1]{\endgroup\@href {#1}{\urlprefix }}%
\providecommand \urlprefix  [0]{URL }%
\providecommand \Eprint [0]{\href }%
\providecommand \doibase [0]{https://doi.org/}%
\providecommand \selectlanguage [0]{\@gobble}%
\providecommand \bibinfo  [0]{\@secondoftwo}%
\providecommand \bibfield  [0]{\@secondoftwo}%
\providecommand \translation [1]{[#1]}%
\providecommand \BibitemOpen [0]{}%
\providecommand \bibitemStop [0]{}%
\providecommand \bibitemNoStop [0]{.\EOS\space}%
\providecommand \EOS [0]{\spacefactor3000\relax}%
\providecommand \BibitemShut  [1]{\csname bibitem#1\endcsname}%
\let\auto@bib@innerbib\@empty
\bibitem [{\citenamefont {Georgescu}\ \emph {et~al.}(2014)\citenamefont
  {Georgescu}, \citenamefont {Ashhab},\ and\ \citenamefont
  {Nori}}]{Georgescu:2014um}%
  \BibitemOpen
  \bibfield  {author} {\bibinfo {author} {\bibfnamefont {I.~M.}\ \bibnamefont
  {Georgescu}}, \bibinfo {author} {\bibfnamefont {S.}~\bibnamefont {Ashhab}},\
  and\ \bibinfo {author} {\bibfnamefont {F.}~\bibnamefont {Nori}},\ }\bibfield
  {title} {\bibinfo {title} {Quantum simulation},\ }\href
  {https://doi.org/10.1103/RevModPhys.86.153} {\bibfield  {journal} {\bibinfo
  {journal} {Rev. Mod. Phys.}\ }\textbf {\bibinfo {volume} {86}},\ \bibinfo
  {pages} {153} (\bibinfo {year} {2014})}\BibitemShut {NoStop}%
\bibitem [{\citenamefont {Abrams}\ and\ \citenamefont
  {Lloyd}(1999)}]{Abrams:1999ur}%
  \BibitemOpen
  \bibfield  {author} {\bibinfo {author} {\bibfnamefont {D.~S.}\ \bibnamefont
  {Abrams}}\ and\ \bibinfo {author} {\bibfnamefont {S.}~\bibnamefont {Lloyd}},\
  }\bibfield  {title} {\bibinfo {title} {Quantum algorithm providing
  exponential speed increase for finding eigenvalues and eigenvectors},\ }\href
  {https://doi.org/10.1103/PhysRevLett.83.5162} {\bibfield  {journal} {\bibinfo
   {journal} {Phys. Rev. Lett.}\ }\textbf {\bibinfo {volume} {83}},\ \bibinfo
  {pages} {5162} (\bibinfo {year} {1999})}\BibitemShut {NoStop}%
\bibitem [{\citenamefont {Parrish}\ and\ \citenamefont
  {McMahon}(2019)}]{parrish2019quantum}%
  \BibitemOpen
  \bibfield  {author} {\bibinfo {author} {\bibfnamefont {R.~M.}\ \bibnamefont
  {Parrish}}\ and\ \bibinfo {author} {\bibfnamefont {P.~L.}\ \bibnamefont
  {McMahon}},\ }\bibfield  {title} {\bibinfo {title} {Quantum filter
  diagonalization: Quantum eigendecomposition without full quantum phase
  estimation},\ }\href@noop {} {\bibfield  {journal} {\bibinfo  {journal}
  {arXiv preprint arXiv:1909.08925}\ } (\bibinfo {year} {2019})}\BibitemShut
  {NoStop}%
\bibitem [{\citenamefont {Stair}\ \emph {et~al.}(2020)\citenamefont {Stair},
  \citenamefont {Huang},\ and\ \citenamefont {Evangelista}}]{stair:Krylov2020}%
  \BibitemOpen
  \bibfield  {author} {\bibinfo {author} {\bibfnamefont {N.~H.}\ \bibnamefont
  {Stair}}, \bibinfo {author} {\bibfnamefont {R.}~\bibnamefont {Huang}},\ and\
  \bibinfo {author} {\bibfnamefont {F.~A.}\ \bibnamefont {Evangelista}},\
  }\bibfield  {title} {\bibinfo {title} {A multireference quantum krylov
  algorithm for strongly correlated electrons},\ }\href@noop {} {\bibfield
  {journal} {\bibinfo  {journal} {J. Chem. Theory Comput.}\ }\textbf {\bibinfo
  {volume} {16}},\ \bibinfo {pages} {2236} (\bibinfo {year}
  {2020})}\BibitemShut {NoStop}%
\bibitem [{\citenamefont {Klymko}\ \emph {et~al.}(2022)\citenamefont {Klymko},
  \citenamefont {Mejuto-Zaera}, \citenamefont {Cotton}, \citenamefont
  {Wudarski}, \citenamefont {Urbanek}, \citenamefont {Hait}, \citenamefont
  {Head-Gordon}, \citenamefont {Whaley}, \citenamefont {Moussa}, \citenamefont
  {Wiebe}, \citenamefont {de~Jong},\ and\ \citenamefont
  {Tubman}}]{Klymko:2022vp}%
  \BibitemOpen
  \bibfield  {author} {\bibinfo {author} {\bibfnamefont {K.}~\bibnamefont
  {Klymko}}, \bibinfo {author} {\bibfnamefont {C.}~\bibnamefont
  {Mejuto-Zaera}}, \bibinfo {author} {\bibfnamefont {S.~J.}\ \bibnamefont
  {Cotton}}, \bibinfo {author} {\bibfnamefont {F.}~\bibnamefont {Wudarski}},
  \bibinfo {author} {\bibfnamefont {M.}~\bibnamefont {Urbanek}}, \bibinfo
  {author} {\bibfnamefont {D.}~\bibnamefont {Hait}}, \bibinfo {author}
  {\bibfnamefont {M.}~\bibnamefont {Head-Gordon}}, \bibinfo {author}
  {\bibfnamefont {K.~B.}\ \bibnamefont {Whaley}}, \bibinfo {author}
  {\bibfnamefont {J.}~\bibnamefont {Moussa}}, \bibinfo {author} {\bibfnamefont
  {N.}~\bibnamefont {Wiebe}}, \bibinfo {author} {\bibfnamefont {W.~A.}\
  \bibnamefont {de~Jong}},\ and\ \bibinfo {author} {\bibfnamefont {N.~M.}\
  \bibnamefont {Tubman}},\ }\bibfield  {title} {\bibinfo {title} {Real-time
  evolution for ultracompact hamiltonian eigenstates on quantum hardware},\
  }\href {https://doi.org/10.1103/PRXQuantum.3.020323} {\bibfield  {journal}
  {\bibinfo  {journal} {PRX Quantum}\ }\textbf {\bibinfo {volume} {3}},\
  \bibinfo {pages} {020323} (\bibinfo {year} {2022})}\BibitemShut {NoStop}%
\bibitem [{\citenamefont {Suzuki}(1990)}]{SUZUKI1990319}%
  \BibitemOpen
  \bibfield  {author} {\bibinfo {author} {\bibfnamefont {M.}~\bibnamefont
  {Suzuki}},\ }\bibfield  {title} {\bibinfo {title} {Fractal decomposition of
  exponential operators with applications to many-body theories and monte carlo
  simulations},\ }\href
  {https://doi.org/https://doi.org/10.1016/0375-9601(90)90962-N} {\bibfield
  {journal} {\bibinfo  {journal} {Phys. Lett. A}\ }\textbf {\bibinfo {volume}
  {146}},\ \bibinfo {pages} {319} (\bibinfo {year} {1990})}\BibitemShut
  {NoStop}%
\bibitem [{\citenamefont {Suzuki}(1991)}]{Suzuki_1991}%
  \BibitemOpen
  \bibfield  {author} {\bibinfo {author} {\bibfnamefont {M.}~\bibnamefont
  {Suzuki}},\ }\bibfield  {title} {\bibinfo {title} {General theory of fractal
  path integrals with applications to many-body theories and statistical
  physics},\ }\href {https://doi.org/10.1063/1.529425} {\bibfield  {journal}
  {\bibinfo  {journal} {J. Math. Phys.}\ }\textbf {\bibinfo {volume} {32}},\
  \bibinfo {pages} {400} (\bibinfo {year} {1991})}\BibitemShut {NoStop}%
\bibitem [{\citenamefont {McClean}\ \emph {et~al.}(2014)\citenamefont
  {McClean}, \citenamefont {Babbush}, \citenamefont {Love},\ and\ \citenamefont
  {Aspuru-Guzik}}]{mcclean2014exploiting}%
  \BibitemOpen
  \bibfield  {author} {\bibinfo {author} {\bibfnamefont {J.~R.}\ \bibnamefont
  {McClean}}, \bibinfo {author} {\bibfnamefont {R.}~\bibnamefont {Babbush}},
  \bibinfo {author} {\bibfnamefont {P.~J.}\ \bibnamefont {Love}},\ and\
  \bibinfo {author} {\bibfnamefont {A.}~\bibnamefont {Aspuru-Guzik}},\
  }\bibfield  {title} {\bibinfo {title} {Exploiting locality in quantum
  computation for quantum chemistry},\ }\href@noop {} {\bibfield  {journal}
  {\bibinfo  {journal} {J. Phys. Chem. Lett.}\ }\textbf {\bibinfo {volume}
  {5}},\ \bibinfo {pages} {4368} (\bibinfo {year} {2014})}\BibitemShut
  {NoStop}%
\bibitem [{\citenamefont {Poulin}\ \emph {et~al.}(2015)\citenamefont {Poulin},
  \citenamefont {Hastings}, \citenamefont {Wecker}, \citenamefont {Wiebe},
  \citenamefont {Doberty},\ and\ \citenamefont {Troyer}}]{poulin2014trotter}%
  \BibitemOpen
  \bibfield  {author} {\bibinfo {author} {\bibfnamefont {D.}~\bibnamefont
  {Poulin}}, \bibinfo {author} {\bibfnamefont {M.~B.}\ \bibnamefont
  {Hastings}}, \bibinfo {author} {\bibfnamefont {D.}~\bibnamefont {Wecker}},
  \bibinfo {author} {\bibfnamefont {N.}~\bibnamefont {Wiebe}}, \bibinfo
  {author} {\bibfnamefont {A.~C.}\ \bibnamefont {Doberty}},\ and\ \bibinfo
  {author} {\bibfnamefont {M.}~\bibnamefont {Troyer}},\ }\bibfield  {title}
  {\bibinfo {title} {The trotter step size required for accurate quantum
  simulation of quantum chemistry},\ }\href
  {https://dl.acm.org/doi/10.5555/2871401.2871402} {\bibfield  {journal}
  {\bibinfo  {journal} {Quant. Info. Comput}\ }\textbf {\bibinfo {volume}
  {15}},\ \bibinfo {pages} {361–384} (\bibinfo {year} {2015})}\BibitemShut
  {NoStop}%
\bibitem [{\citenamefont {Motta}\ \emph {et~al.}(2021)\citenamefont {Motta},
  \citenamefont {Ye}, \citenamefont {McClean}, \citenamefont {Li},
  \citenamefont {Minnich}, \citenamefont {Babbush},\ and\ \citenamefont
  {Chan}}]{Motta_2021}%
  \BibitemOpen
  \bibfield  {author} {\bibinfo {author} {\bibfnamefont {M.}~\bibnamefont
  {Motta}}, \bibinfo {author} {\bibfnamefont {E.}~\bibnamefont {Ye}}, \bibinfo
  {author} {\bibfnamefont {J.~R.}\ \bibnamefont {McClean}}, \bibinfo {author}
  {\bibfnamefont {Z.}~\bibnamefont {Li}}, \bibinfo {author} {\bibfnamefont
  {A.~J.}\ \bibnamefont {Minnich}}, \bibinfo {author} {\bibfnamefont
  {R.}~\bibnamefont {Babbush}},\ and\ \bibinfo {author} {\bibfnamefont
  {G.~K.-L.}\ \bibnamefont {Chan}},\ }\bibfield  {title} {\bibinfo {title} {Low
  rank representations for quantum simulation of electronic structure},\ }\href
  {https://doi.org/10.1038%2Fs41534-021-00416-z} {\bibfield  {journal}
  {\bibinfo  {journal} {Npj Quantum Inf.}\ }\textbf {\bibinfo {volume} {7}}
  (\bibinfo {year} {2021})}\BibitemShut {NoStop}%
\bibitem [{\citenamefont {Kivlichan}\ \emph {et~al.}(2018)\citenamefont
  {Kivlichan}, \citenamefont {McClean}, \citenamefont {Wiebe}, \citenamefont
  {Gidney}, \citenamefont {Aspuru-Guzik}, \citenamefont {Chan},\ and\
  \citenamefont {Babbush}}]{Kivlichan:2021ld}%
  \BibitemOpen
  \bibfield  {author} {\bibinfo {author} {\bibfnamefont {I.~D.}\ \bibnamefont
  {Kivlichan}}, \bibinfo {author} {\bibfnamefont {J.}~\bibnamefont {McClean}},
  \bibinfo {author} {\bibfnamefont {N.}~\bibnamefont {Wiebe}}, \bibinfo
  {author} {\bibfnamefont {C.}~\bibnamefont {Gidney}}, \bibinfo {author}
  {\bibfnamefont {A.}~\bibnamefont {Aspuru-Guzik}}, \bibinfo {author}
  {\bibfnamefont {G.~K.-L.}\ \bibnamefont {Chan}},\ and\ \bibinfo {author}
  {\bibfnamefont {R.}~\bibnamefont {Babbush}},\ }\bibfield  {title} {\bibinfo
  {title} {Quantum simulation of electronic structure with linear depth and
  connectivity},\ }\href {https://doi.org/10.1103/PhysRevLett.120.110501}
  {\bibfield  {journal} {\bibinfo  {journal} {Phys. Rev. Lett.}\ }\textbf
  {\bibinfo {volume} {120}},\ \bibinfo {pages} {110501} (\bibinfo {year}
  {2018})}\BibitemShut {NoStop}%
\bibitem [{\citenamefont {Huggins}\ \emph {et~al.}(2021)\citenamefont
  {Huggins}, \citenamefont {McClean}, \citenamefont {Rubin}, \citenamefont
  {Jiang}, \citenamefont {Wiebe}, \citenamefont {Whaley},\ and\ \citenamefont
  {Babbush}}]{huggins2021efficient}%
  \BibitemOpen
  \bibfield  {author} {\bibinfo {author} {\bibfnamefont {W.~J.}\ \bibnamefont
  {Huggins}}, \bibinfo {author} {\bibfnamefont {J.~R.}\ \bibnamefont
  {McClean}}, \bibinfo {author} {\bibfnamefont {N.~C.}\ \bibnamefont {Rubin}},
  \bibinfo {author} {\bibfnamefont {Z.}~\bibnamefont {Jiang}}, \bibinfo
  {author} {\bibfnamefont {N.}~\bibnamefont {Wiebe}}, \bibinfo {author}
  {\bibfnamefont {K.~B.}\ \bibnamefont {Whaley}},\ and\ \bibinfo {author}
  {\bibfnamefont {R.}~\bibnamefont {Babbush}},\ }\bibfield  {title} {\bibinfo
  {title} {Efficient and noise resilient measurements for quantum chemistry on
  near-term quantum computers},\ }\href@noop {} {\bibfield  {journal} {\bibinfo
   {journal} {Npj Quantum Inf.}\ }\textbf {\bibinfo {volume} {7}},\ \bibinfo
  {pages} {1} (\bibinfo {year} {2021})}\BibitemShut {NoStop}%
\bibitem [{\citenamefont {Childs}\ \emph {et~al.}(2019)\citenamefont {Childs},
  \citenamefont {Ostrander},\ and\ \citenamefont {Su}}]{childs2019faster}%
  \BibitemOpen
  \bibfield  {author} {\bibinfo {author} {\bibfnamefont {A.~M.}\ \bibnamefont
  {Childs}}, \bibinfo {author} {\bibfnamefont {A.}~\bibnamefont {Ostrander}},\
  and\ \bibinfo {author} {\bibfnamefont {Y.}~\bibnamefont {Su}},\ }\bibfield
  {title} {\bibinfo {title} {Faster quantum simulation by randomization},\
  }\href@noop {} {\bibfield  {journal} {\bibinfo  {journal} {Quantum}\ }\textbf
  {\bibinfo {volume} {3}},\ \bibinfo {pages} {182} (\bibinfo {year}
  {2019})}\BibitemShut {NoStop}%
\bibitem [{\citenamefont {Campbell}(2019)}]{Campbell:2019qdrift}%
  \BibitemOpen
  \bibfield  {author} {\bibinfo {author} {\bibfnamefont {E.}~\bibnamefont
  {Campbell}},\ }\bibfield  {title} {\bibinfo {title} {Random compiler for fast
  hamiltonian simulation},\ }\href
  {https://doi.org/10.1103/PhysRevLett.123.070503} {\bibfield  {journal}
  {\bibinfo  {journal} {Phys. Rev. Lett.}\ }\textbf {\bibinfo {volume} {123}},\
  \bibinfo {pages} {070503} (\bibinfo {year} {2019})}\BibitemShut {NoStop}%
\bibitem [{\citenamefont {Kivlichan}\ \emph {et~al.}(2019)\citenamefont
  {Kivlichan}, \citenamefont {Granade},\ and\ \citenamefont
  {Wiebe}}]{kivlichan2019phase}%
  \BibitemOpen
  \bibfield  {author} {\bibinfo {author} {\bibfnamefont {I.~D.}\ \bibnamefont
  {Kivlichan}}, \bibinfo {author} {\bibfnamefont {C.~E.}\ \bibnamefont
  {Granade}},\ and\ \bibinfo {author} {\bibfnamefont {N.}~\bibnamefont
  {Wiebe}},\ }\bibfield  {title} {\bibinfo {title} {Phase estimation with
  randomized hamiltonians},\ }\href@noop {} {\bibfield  {journal} {\bibinfo
  {journal} {arXiv preprint arXiv:1907.10070}\ } (\bibinfo {year}
  {2019})}\BibitemShut {NoStop}%
\bibitem [{\citenamefont {Ouyang}\ \emph {et~al.}(2020)\citenamefont {Ouyang},
  \citenamefont {White},\ and\ \citenamefont
  {Campbell}}]{ouyang2020compilation}%
  \BibitemOpen
  \bibfield  {author} {\bibinfo {author} {\bibfnamefont {Y.}~\bibnamefont
  {Ouyang}}, \bibinfo {author} {\bibfnamefont {D.~R.}\ \bibnamefont {White}},\
  and\ \bibinfo {author} {\bibfnamefont {E.~T.}\ \bibnamefont {Campbell}},\
  }\bibfield  {title} {\bibinfo {title} {Compilation by stochastic hamiltonian
  sparsification},\ }\href@noop {} {\bibfield  {journal} {\bibinfo  {journal}
  {Quantum}\ }\textbf {\bibinfo {volume} {4}},\ \bibinfo {pages} {235}
  (\bibinfo {year} {2020})}\BibitemShut {NoStop}%
\bibitem [{\citenamefont {Wan}\ \emph {et~al.}(2022)\citenamefont {Wan},
  \citenamefont {Berta},\ and\ \citenamefont {Campbell}}]{wan2022randomized}%
  \BibitemOpen
  \bibfield  {author} {\bibinfo {author} {\bibfnamefont {K.}~\bibnamefont
  {Wan}}, \bibinfo {author} {\bibfnamefont {M.}~\bibnamefont {Berta}},\ and\
  \bibinfo {author} {\bibfnamefont {E.~T.}\ \bibnamefont {Campbell}},\
  }\bibfield  {title} {\bibinfo {title} {Randomized quantum algorithm for
  statistical phase estimation},\ }\href@noop {} {\bibfield  {journal}
  {\bibinfo  {journal} {Phys. Rev. Lett.}\ }\textbf {\bibinfo {volume} {129}},\
  \bibinfo {pages} {030503} (\bibinfo {year} {2022})}\BibitemShut {NoStop}%
\bibitem [{\citenamefont {Childs}\ and\ \citenamefont
  {Wiebe}(2012)}]{childs:2012lcu}%
  \BibitemOpen
  \bibfield  {author} {\bibinfo {author} {\bibfnamefont {A.~M.}\ \bibnamefont
  {Childs}}\ and\ \bibinfo {author} {\bibfnamefont {N.}~\bibnamefont {Wiebe}},\
  }\bibfield  {title} {\bibinfo {title} {Hamiltonian simulation using linear
  combinations of unitary operations},\ }\href@noop {} {\bibfield  {journal}
  {\bibinfo  {journal} {Quantum Info. Comput.}\ }\textbf {\bibinfo {volume}
  {12}},\ \bibinfo {pages} {901} (\bibinfo {year} {2012})}\BibitemShut
  {NoStop}%
\bibitem [{\citenamefont {Babbush}\ \emph {et~al.}(2016)\citenamefont
  {Babbush}, \citenamefont {Berry}, \citenamefont {Kivlichan}, \citenamefont
  {Wei}, \citenamefont {Love},\ and\ \citenamefont
  {Aspuru-Guzik}}]{Babbush_2016}%
  \BibitemOpen
  \bibfield  {author} {\bibinfo {author} {\bibfnamefont {R.}~\bibnamefont
  {Babbush}}, \bibinfo {author} {\bibfnamefont {D.~W.}\ \bibnamefont {Berry}},
  \bibinfo {author} {\bibfnamefont {I.~D.}\ \bibnamefont {Kivlichan}}, \bibinfo
  {author} {\bibfnamefont {A.~Y.}\ \bibnamefont {Wei}}, \bibinfo {author}
  {\bibfnamefont {P.~J.}\ \bibnamefont {Love}},\ and\ \bibinfo {author}
  {\bibfnamefont {A.}~\bibnamefont {Aspuru-Guzik}},\ }\bibfield  {title}
  {\bibinfo {title} {Exponentially more precise quantum simulation of fermions
  in second quantization},\ }\href
  {https://doi.org/10.1088/1367-2630/18/3/033032} {\bibfield  {journal}
  {\bibinfo  {journal} {New J. Phys.}\ }\textbf {\bibinfo {volume} {18}},\
  \bibinfo {pages} {033032} (\bibinfo {year} {2016})}\BibitemShut {NoStop}%
\bibitem [{\citenamefont {Low}\ and\ \citenamefont
  {Chuang}(2017)}]{low:2017qsp}%
  \BibitemOpen
  \bibfield  {author} {\bibinfo {author} {\bibfnamefont {G.~H.}\ \bibnamefont
  {Low}}\ and\ \bibinfo {author} {\bibfnamefont {I.~L.}\ \bibnamefont
  {Chuang}},\ }\bibfield  {title} {\bibinfo {title} {Optimal hamiltonian
  simulation by quantum signal processing},\ }\href
  {https://doi.org/10.1103/PhysRevLett.118.010501} {\bibfield  {journal}
  {\bibinfo  {journal} {Phys. Rev. Lett.}\ }\textbf {\bibinfo {volume} {118}},\
  \bibinfo {pages} {010501} (\bibinfo {year} {2017})}\BibitemShut {NoStop}%
\bibitem [{\citenamefont {Berry}\ \emph {et~al.}(2019)\citenamefont {Berry},
  \citenamefont {Gidney}, \citenamefont {Motta}, \citenamefont {McClean},\ and\
  \citenamefont {Babbush}}]{Berry2019qubitizationof}%
  \BibitemOpen
  \bibfield  {author} {\bibinfo {author} {\bibfnamefont {D.~W.}\ \bibnamefont
  {Berry}}, \bibinfo {author} {\bibfnamefont {C.}~\bibnamefont {Gidney}},
  \bibinfo {author} {\bibfnamefont {M.}~\bibnamefont {Motta}}, \bibinfo
  {author} {\bibfnamefont {J.~R.}\ \bibnamefont {McClean}},\ and\ \bibinfo
  {author} {\bibfnamefont {R.}~\bibnamefont {Babbush}},\ }\bibfield  {title}
  {\bibinfo {title} {Qubitization of {A}rbitrary {B}asis {Q}uantum {C}hemistry
  {L}everaging {S}parsity and {L}ow {R}ank {F}actorization},\ }\href
  {https://doi.org/10.22331/q-2019-12-02-208} {\bibfield  {journal} {\bibinfo
  {journal} {{Quantum}}\ }\textbf {\bibinfo {volume} {3}},\ \bibinfo {pages}
  {208} (\bibinfo {year} {2019})}\BibitemShut {NoStop}%
\bibitem [{\citenamefont {McClean}\ \emph {et~al.}(2017)\citenamefont
  {McClean}, \citenamefont {Kimchi-Schwartz}, \citenamefont {Carter},\ and\
  \citenamefont {de~Jong}}]{mcclean:2017qse}%
  \BibitemOpen
  \bibfield  {author} {\bibinfo {author} {\bibfnamefont {J.~R.}\ \bibnamefont
  {McClean}}, \bibinfo {author} {\bibfnamefont {M.~E.}\ \bibnamefont
  {Kimchi-Schwartz}}, \bibinfo {author} {\bibfnamefont {J.}~\bibnamefont
  {Carter}},\ and\ \bibinfo {author} {\bibfnamefont {W.~A.}\ \bibnamefont
  {de~Jong}},\ }\bibfield  {title} {\bibinfo {title} {Hybrid quantum-classical
  hierarchy for mitigation of decoherence and determination of excited
  states},\ }\href {https://doi.org/10.1103/PhysRevA.95.042308} {\bibfield
  {journal} {\bibinfo  {journal} {Phys. Rev. A}\ }\textbf {\bibinfo {volume}
  {95}},\ \bibinfo {pages} {042308} (\bibinfo {year} {2017})}\BibitemShut
  {NoStop}%
\bibitem [{\citenamefont {Motta}\ \emph {et~al.}(2020)\citenamefont {Motta},
  \citenamefont {Sun}, \citenamefont {Tan}, \citenamefont {O'Rourke},
  \citenamefont {Ye}, \citenamefont {Minnich}, \citenamefont {Brandao},\ and\
  \citenamefont {Chan}}]{motta2020determining}%
  \BibitemOpen
  \bibfield  {author} {\bibinfo {author} {\bibfnamefont {M.}~\bibnamefont
  {Motta}}, \bibinfo {author} {\bibfnamefont {C.}~\bibnamefont {Sun}}, \bibinfo
  {author} {\bibfnamefont {A.~T.}\ \bibnamefont {Tan}}, \bibinfo {author}
  {\bibfnamefont {M.~J.}\ \bibnamefont {O'Rourke}}, \bibinfo {author}
  {\bibfnamefont {E.}~\bibnamefont {Ye}}, \bibinfo {author} {\bibfnamefont
  {A.~J.}\ \bibnamefont {Minnich}}, \bibinfo {author} {\bibfnamefont {F.~G.}\
  \bibnamefont {Brandao}},\ and\ \bibinfo {author} {\bibfnamefont {G.~K.}\
  \bibnamefont {Chan}},\ }\bibfield  {title} {\bibinfo {title} {Determining
  eigenstates and thermal states on a quantum computer using quantum imaginary
  time evolution},\ }\href@noop {} {\bibfield  {journal} {\bibinfo  {journal}
  {Nat. Phys.}\ }\textbf {\bibinfo {volume} {16}},\ \bibinfo {pages} {205}
  (\bibinfo {year} {2020})}\BibitemShut {NoStop}%
\bibitem [{\citenamefont {McClean}\ \emph {et~al.}(2020)\citenamefont
  {McClean}, \citenamefont {Jiang}, \citenamefont {Rubin}, \citenamefont
  {Babbush},\ and\ \citenamefont {Neven}}]{mcclean2020decoding}%
  \BibitemOpen
  \bibfield  {author} {\bibinfo {author} {\bibfnamefont {J.~R.}\ \bibnamefont
  {McClean}}, \bibinfo {author} {\bibfnamefont {Z.}~\bibnamefont {Jiang}},
  \bibinfo {author} {\bibfnamefont {N.~C.}\ \bibnamefont {Rubin}}, \bibinfo
  {author} {\bibfnamefont {R.}~\bibnamefont {Babbush}},\ and\ \bibinfo {author}
  {\bibfnamefont {H.}~\bibnamefont {Neven}},\ }\bibfield  {title} {\bibinfo
  {title} {Decoding quantum errors with subspace expansions},\ }\href@noop {}
  {\bibfield  {journal} {\bibinfo  {journal} {Nat. Commun.}\ }\textbf {\bibinfo
  {volume} {11}},\ \bibinfo {pages} {1} (\bibinfo {year} {2020})}\BibitemShut
  {NoStop}%
\bibitem [{\citenamefont {Takeshita}\ \emph {et~al.}(2020)\citenamefont
  {Takeshita}, \citenamefont {Rubin}, \citenamefont {Jiang}, \citenamefont
  {Lee}, \citenamefont {Babbush},\ and\ \citenamefont
  {McClean}}]{takeshita2020increasing}%
  \BibitemOpen
  \bibfield  {author} {\bibinfo {author} {\bibfnamefont {T.}~\bibnamefont
  {Takeshita}}, \bibinfo {author} {\bibfnamefont {N.~C.}\ \bibnamefont
  {Rubin}}, \bibinfo {author} {\bibfnamefont {Z.}~\bibnamefont {Jiang}},
  \bibinfo {author} {\bibfnamefont {E.}~\bibnamefont {Lee}}, \bibinfo {author}
  {\bibfnamefont {R.}~\bibnamefont {Babbush}},\ and\ \bibinfo {author}
  {\bibfnamefont {J.~R.}\ \bibnamefont {McClean}},\ }\bibfield  {title}
  {\bibinfo {title} {Increasing the representation accuracy of quantum
  simulations of chemistry without extra quantum resources},\ }\href@noop {}
  {\bibfield  {journal} {\bibinfo  {journal} {Phys. Rev. X}\ }\textbf {\bibinfo
  {volume} {10}},\ \bibinfo {pages} {011004} (\bibinfo {year}
  {2020})}\BibitemShut {NoStop}%
\bibitem [{\citenamefont {Yoshioka}\ \emph {et~al.}(2022)\citenamefont
  {Yoshioka}, \citenamefont {Hakoshima}, \citenamefont {Matsuzaki},
  \citenamefont {Tokunaga}, \citenamefont {Suzuki},\ and\ \citenamefont
  {Endo}}]{yoshioka2022generalized}%
  \BibitemOpen
  \bibfield  {author} {\bibinfo {author} {\bibfnamefont {N.}~\bibnamefont
  {Yoshioka}}, \bibinfo {author} {\bibfnamefont {H.}~\bibnamefont {Hakoshima}},
  \bibinfo {author} {\bibfnamefont {Y.}~\bibnamefont {Matsuzaki}}, \bibinfo
  {author} {\bibfnamefont {Y.}~\bibnamefont {Tokunaga}}, \bibinfo {author}
  {\bibfnamefont {Y.}~\bibnamefont {Suzuki}},\ and\ \bibinfo {author}
  {\bibfnamefont {S.}~\bibnamefont {Endo}},\ }\bibfield  {title} {\bibinfo
  {title} {Generalized quantum subspace expansion},\ }\href@noop {} {\bibfield
  {journal} {\bibinfo  {journal} {Phys. Rev. Lett.}\ }\textbf {\bibinfo
  {volume} {129}},\ \bibinfo {pages} {020502} (\bibinfo {year}
  {2022})}\BibitemShut {NoStop}%
\bibitem [{\citenamefont {Ollitrault}\ \emph {et~al.}(2020)\citenamefont
  {Ollitrault}, \citenamefont {Kandala}, \citenamefont {Chen}, \citenamefont
  {Barkoutsos}, \citenamefont {Mezzacapo}, \citenamefont {Pistoia},
  \citenamefont {Sheldon}, \citenamefont {Woerner}, \citenamefont {Gambetta},\
  and\ \citenamefont {Tavernelli}}]{ollitrault2020quantum}%
  \BibitemOpen
  \bibfield  {author} {\bibinfo {author} {\bibfnamefont {P.~J.}\ \bibnamefont
  {Ollitrault}}, \bibinfo {author} {\bibfnamefont {A.}~\bibnamefont {Kandala}},
  \bibinfo {author} {\bibfnamefont {C.-F.}\ \bibnamefont {Chen}}, \bibinfo
  {author} {\bibfnamefont {P.~K.}\ \bibnamefont {Barkoutsos}}, \bibinfo
  {author} {\bibfnamefont {A.}~\bibnamefont {Mezzacapo}}, \bibinfo {author}
  {\bibfnamefont {M.}~\bibnamefont {Pistoia}}, \bibinfo {author} {\bibfnamefont
  {S.}~\bibnamefont {Sheldon}}, \bibinfo {author} {\bibfnamefont
  {S.}~\bibnamefont {Woerner}}, \bibinfo {author} {\bibfnamefont {J.~M.}\
  \bibnamefont {Gambetta}},\ and\ \bibinfo {author} {\bibfnamefont
  {I.}~\bibnamefont {Tavernelli}},\ }\bibfield  {title} {\bibinfo {title}
  {Quantum equation of motion for computing molecular excitation energies on a
  noisy quantum processor},\ }\href@noop {} {\bibfield  {journal} {\bibinfo
  {journal} {Phys. Rev. Res.}\ }\textbf {\bibinfo {volume} {2}},\ \bibinfo
  {pages} {043140} (\bibinfo {year} {2020})}\BibitemShut {NoStop}%
\bibitem [{\citenamefont {Ganzhorn}\ \emph {et~al.}(2019)\citenamefont
  {Ganzhorn}, \citenamefont {Egger}, \citenamefont {Barkoutsos}, \citenamefont
  {Ollitrault}, \citenamefont {Salis}, \citenamefont {Moll}, \citenamefont
  {Roth}, \citenamefont {Fuhrer}, \citenamefont {Mueller}, \citenamefont
  {Woerner} \emph {et~al.}}]{ganzhorn2019gate}%
  \BibitemOpen
  \bibfield  {author} {\bibinfo {author} {\bibfnamefont {M.}~\bibnamefont
  {Ganzhorn}}, \bibinfo {author} {\bibfnamefont {D.~J.}\ \bibnamefont {Egger}},
  \bibinfo {author} {\bibfnamefont {P.}~\bibnamefont {Barkoutsos}}, \bibinfo
  {author} {\bibfnamefont {P.}~\bibnamefont {Ollitrault}}, \bibinfo {author}
  {\bibfnamefont {G.}~\bibnamefont {Salis}}, \bibinfo {author} {\bibfnamefont
  {N.}~\bibnamefont {Moll}}, \bibinfo {author} {\bibfnamefont {M.}~\bibnamefont
  {Roth}}, \bibinfo {author} {\bibfnamefont {A.}~\bibnamefont {Fuhrer}},
  \bibinfo {author} {\bibfnamefont {P.}~\bibnamefont {Mueller}}, \bibinfo
  {author} {\bibfnamefont {S.}~\bibnamefont {Woerner}}, \emph {et~al.},\
  }\bibfield  {title} {\bibinfo {title} {Gate-efficient simulation of molecular
  eigenstates on a quantum computer},\ }\href@noop {} {\bibfield  {journal}
  {\bibinfo  {journal} {Phys. Lett. Appl.}\ }\textbf {\bibinfo {volume} {11}},\
  \bibinfo {pages} {044092} (\bibinfo {year} {2019})}\BibitemShut {NoStop}%
\bibitem [{\citenamefont {Gao}\ \emph {et~al.}(2021)\citenamefont {Gao},
  \citenamefont {Jones}, \citenamefont {Motta}, \citenamefont {Sugawara},
  \citenamefont {Watanabe}, \citenamefont {Kobayashi}, \citenamefont
  {Watanabe}, \citenamefont {Ohnishi}, \citenamefont {Nakamura},\ and\
  \citenamefont {Yamamoto}}]{gao2021applications}%
  \BibitemOpen
  \bibfield  {author} {\bibinfo {author} {\bibfnamefont {Q.}~\bibnamefont
  {Gao}}, \bibinfo {author} {\bibfnamefont {G.~O.}\ \bibnamefont {Jones}},
  \bibinfo {author} {\bibfnamefont {M.}~\bibnamefont {Motta}}, \bibinfo
  {author} {\bibfnamefont {M.}~\bibnamefont {Sugawara}}, \bibinfo {author}
  {\bibfnamefont {H.~C.}\ \bibnamefont {Watanabe}}, \bibinfo {author}
  {\bibfnamefont {T.}~\bibnamefont {Kobayashi}}, \bibinfo {author}
  {\bibfnamefont {E.}~\bibnamefont {Watanabe}}, \bibinfo {author}
  {\bibfnamefont {Y.-y.}\ \bibnamefont {Ohnishi}}, \bibinfo {author}
  {\bibfnamefont {H.}~\bibnamefont {Nakamura}},\ and\ \bibinfo {author}
  {\bibfnamefont {N.}~\bibnamefont {Yamamoto}},\ }\bibfield  {title} {\bibinfo
  {title} {Applications of quantum computing for investigations of electronic
  transitions in phenylsulfonyl-carbazole tadf emitters},\ }\href@noop {}
  {\bibfield  {journal} {\bibinfo  {journal} {Npj Comput. Mater.}\ }\textbf
  {\bibinfo {volume} {7}},\ \bibinfo {pages} {1} (\bibinfo {year}
  {2021})}\BibitemShut {NoStop}%
\bibitem [{\citenamefont {Barison}\ \emph {et~al.}(2022)\citenamefont
  {Barison}, \citenamefont {Galli},\ and\ \citenamefont
  {Motta}}]{barison2022quantum}%
  \BibitemOpen
  \bibfield  {author} {\bibinfo {author} {\bibfnamefont {S.}~\bibnamefont
  {Barison}}, \bibinfo {author} {\bibfnamefont {D.~E.}\ \bibnamefont {Galli}},\
  and\ \bibinfo {author} {\bibfnamefont {M.}~\bibnamefont {Motta}},\ }\bibfield
   {title} {\bibinfo {title} {Quantum simulations of molecular systems with
  intrinsic atomic orbitals},\ }\href@noop {} {\bibfield  {journal} {\bibinfo
  {journal} {Phys. Rev. A}\ }\textbf {\bibinfo {volume} {106}},\ \bibinfo
  {pages} {022404} (\bibinfo {year} {2022})}\BibitemShut {NoStop}%
\bibitem [{\citenamefont {Huggins}\ \emph {et~al.}(2020)\citenamefont
  {Huggins}, \citenamefont {Lee}, \citenamefont {Baek}, \citenamefont
  {O'Gorman},\ and\ \citenamefont {Whaley}}]{huggins2020non}%
  \BibitemOpen
  \bibfield  {author} {\bibinfo {author} {\bibfnamefont {W.~J.}\ \bibnamefont
  {Huggins}}, \bibinfo {author} {\bibfnamefont {J.}~\bibnamefont {Lee}},
  \bibinfo {author} {\bibfnamefont {U.}~\bibnamefont {Baek}}, \bibinfo {author}
  {\bibfnamefont {B.}~\bibnamefont {O'Gorman}},\ and\ \bibinfo {author}
  {\bibfnamefont {K.~B.}\ \bibnamefont {Whaley}},\ }\bibfield  {title}
  {\bibinfo {title} {A non-orthogonal variational quantum eigensolver},\
  }\href@noop {} {\bibfield  {journal} {\bibinfo  {journal} {New J. Phys.}\
  }\textbf {\bibinfo {volume} {22}},\ \bibinfo {pages} {073009} (\bibinfo
  {year} {2020})}\BibitemShut {NoStop}%
\bibitem [{\citenamefont {Baek}\ \emph {et~al.}(2022)\citenamefont {Baek},
  \citenamefont {Hait}, \citenamefont {Shee}, \citenamefont {Leimkuhler},
  \citenamefont {Huggins}, \citenamefont {Stetina}, \citenamefont
  {Head-Gordon},\ and\ \citenamefont {Whaley}}]{baek2022say}%
  \BibitemOpen
  \bibfield  {author} {\bibinfo {author} {\bibfnamefont {U.}~\bibnamefont
  {Baek}}, \bibinfo {author} {\bibfnamefont {D.}~\bibnamefont {Hait}}, \bibinfo
  {author} {\bibfnamefont {J.}~\bibnamefont {Shee}}, \bibinfo {author}
  {\bibfnamefont {O.}~\bibnamefont {Leimkuhler}}, \bibinfo {author}
  {\bibfnamefont {W.~J.}\ \bibnamefont {Huggins}}, \bibinfo {author}
  {\bibfnamefont {T.~F.}\ \bibnamefont {Stetina}}, \bibinfo {author}
  {\bibfnamefont {M.}~\bibnamefont {Head-Gordon}},\ and\ \bibinfo {author}
  {\bibfnamefont {K.~B.}\ \bibnamefont {Whaley}},\ }\bibfield  {title}
  {\bibinfo {title} {Say no to optimization: A non-orthogonal quantum
  eigensolver},\ }\href@noop {} {\bibfield  {journal} {\bibinfo  {journal}
  {arXiv preprint arXiv:2205.09039}\ } (\bibinfo {year} {2022})}\BibitemShut
  {NoStop}%
\bibitem [{\citenamefont {Yeter-Aydeniz}\ \emph {et~al.}(2020)\citenamefont
  {Yeter-Aydeniz}, \citenamefont {Pooser},\ and\ \citenamefont
  {Siopsis}}]{yeter2020practical}%
  \BibitemOpen
  \bibfield  {author} {\bibinfo {author} {\bibfnamefont {K.}~\bibnamefont
  {Yeter-Aydeniz}}, \bibinfo {author} {\bibfnamefont {R.~C.}\ \bibnamefont
  {Pooser}},\ and\ \bibinfo {author} {\bibfnamefont {G.}~\bibnamefont
  {Siopsis}},\ }\bibfield  {title} {\bibinfo {title} {Practical quantum
  computation of chemical and nuclear energy levels using quantum imaginary
  time evolution and lanczos algorithms},\ }\href@noop {} {\bibfield  {journal}
  {\bibinfo  {journal} {Npj Quantum Inf.}\ }\textbf {\bibinfo {volume} {6}},\
  \bibinfo {pages} {1} (\bibinfo {year} {2020})}\BibitemShut {NoStop}%
\bibitem [{\citenamefont {Seki}\ and\ \citenamefont
  {Yunoki}(2021)}]{seki:2021qpm}%
  \BibitemOpen
  \bibfield  {author} {\bibinfo {author} {\bibfnamefont {K.}~\bibnamefont
  {Seki}}\ and\ \bibinfo {author} {\bibfnamefont {S.}~\bibnamefont {Yunoki}},\
  }\bibfield  {title} {\bibinfo {title} {Quantum power method by a
  superposition of time-evolved states},\ }\href
  {https://doi.org/10.1103/PRXQuantum.2.010333} {\bibfield  {journal} {\bibinfo
   {journal} {PRX Quantum}\ }\textbf {\bibinfo {volume} {2}},\ \bibinfo {pages}
  {010333} (\bibinfo {year} {2021})}\BibitemShut {NoStop}%
\bibitem [{\citenamefont {Cohn}\ \emph {et~al.}(2021)\citenamefont {Cohn},
  \citenamefont {Motta},\ and\ \citenamefont {Parrish}}]{Cohn:2021cdf}%
  \BibitemOpen
  \bibfield  {author} {\bibinfo {author} {\bibfnamefont {J.}~\bibnamefont
  {Cohn}}, \bibinfo {author} {\bibfnamefont {M.}~\bibnamefont {Motta}},\ and\
  \bibinfo {author} {\bibfnamefont {R.~M.}\ \bibnamefont {Parrish}},\
  }\bibfield  {title} {\bibinfo {title} {Quantum filter diagonalization with
  compressed double-factorized hamiltonians},\ }\href
  {https://doi.org/10.1103/PRXQuantum.2.040352} {\bibfield  {journal} {\bibinfo
   {journal} {PRX Quantum}\ }\textbf {\bibinfo {volume} {2}},\ \bibinfo {pages}
  {040352} (\bibinfo {year} {2021})}\BibitemShut {NoStop}%
\bibitem [{\citenamefont {Cortes}\ and\ \citenamefont
  {Gray}(2022)}]{cortes2022:qk_es}%
  \BibitemOpen
  \bibfield  {author} {\bibinfo {author} {\bibfnamefont {C.~L.}\ \bibnamefont
  {Cortes}}\ and\ \bibinfo {author} {\bibfnamefont {S.~K.}\ \bibnamefont
  {Gray}},\ }\bibfield  {title} {\bibinfo {title} {Quantum krylov subspace
  algorithms for ground- and excited-state energy estimation},\ }\href
  {https://doi.org/10.1103/PhysRevA.105.022417} {\bibfield  {journal} {\bibinfo
   {journal} {Phys. Rev. A}\ }\textbf {\bibinfo {volume} {105}},\ \bibinfo
  {pages} {022417} (\bibinfo {year} {2022})}\BibitemShut {NoStop}%
\bibitem [{\citenamefont {Cortes}\ \emph {et~al.}(2022)\citenamefont {Cortes},
  \citenamefont {DePrince~III},\ and\ \citenamefont {Gray}}]{cortes2022fast}%
  \BibitemOpen
  \bibfield  {author} {\bibinfo {author} {\bibfnamefont {C.~L.}\ \bibnamefont
  {Cortes}}, \bibinfo {author} {\bibfnamefont {A.~E.}\ \bibnamefont
  {DePrince~III}},\ and\ \bibinfo {author} {\bibfnamefont {S.~K.}\ \bibnamefont
  {Gray}},\ }\bibfield  {title} {\bibinfo {title} {Fast-forwarding quantum
  simulation with real-time quantum krylov subspace algorithms},\ }\href@noop
  {} {\bibfield  {journal} {\bibinfo  {journal} {Phys. Rev. A}\ }\textbf
  {\bibinfo {volume} {106}},\ \bibinfo {pages} {042409} (\bibinfo {year}
  {2022})}\BibitemShut {NoStop}%
\bibitem [{\citenamefont {Shen}\ \emph {et~al.}(2022)\citenamefont {Shen},
  \citenamefont {Klymko}, \citenamefont {Sud}, \citenamefont {Williams-Young},
  \citenamefont {de~Jong},\ and\ \citenamefont {Tubman}}]{shen2022real}%
  \BibitemOpen
  \bibfield  {author} {\bibinfo {author} {\bibfnamefont {Y.}~\bibnamefont
  {Shen}}, \bibinfo {author} {\bibfnamefont {K.}~\bibnamefont {Klymko}},
  \bibinfo {author} {\bibfnamefont {J.}~\bibnamefont {Sud}}, \bibinfo {author}
  {\bibfnamefont {D.~B.}\ \bibnamefont {Williams-Young}}, \bibinfo {author}
  {\bibfnamefont {W.~A.}\ \bibnamefont {de~Jong}},\ and\ \bibinfo {author}
  {\bibfnamefont {N.~M.}\ \bibnamefont {Tubman}},\ }\bibfield  {title}
  {\bibinfo {title} {Real-time krylov theory for quantum computing
  algorithms},\ }\href@noop {} {\bibfield  {journal} {\bibinfo  {journal}
  {arXiv preprint arXiv:2208.01063}\ } (\bibinfo {year} {2022})}\BibitemShut
  {NoStop}%
\bibitem [{\citenamefont {Neuhauser}(1990)}]{neuhauser1990bound}%
  \BibitemOpen
  \bibfield  {author} {\bibinfo {author} {\bibfnamefont {D.}~\bibnamefont
  {Neuhauser}},\ }\bibfield  {title} {\bibinfo {title} {Bound state
  eigenfunctions from wave packets: Time→ energy resolution},\ }\href@noop {}
  {\bibfield  {journal} {\bibinfo  {journal} {J. Chem. Phys.}\ }\textbf
  {\bibinfo {volume} {93}},\ \bibinfo {pages} {2611} (\bibinfo {year}
  {1990})}\BibitemShut {NoStop}%
\bibitem [{\citenamefont {Neuhauser}(1994)}]{neuhauser1994circumventing}%
  \BibitemOpen
  \bibfield  {author} {\bibinfo {author} {\bibfnamefont {D.}~\bibnamefont
  {Neuhauser}},\ }\bibfield  {title} {\bibinfo {title} {Circumventing the
  heisenberg principle: A rigorous demonstration of filter-diagonalization on a
  licn model},\ }\href@noop {} {\bibfield  {journal} {\bibinfo  {journal} {J.
  Chem. Phys.}\ }\textbf {\bibinfo {volume} {100}},\ \bibinfo {pages} {5076}
  (\bibinfo {year} {1994})}\BibitemShut {NoStop}%
\bibitem [{\citenamefont {Wall}\ and\ \citenamefont
  {Neuhauser}(1995)}]{wall1995extraction}%
  \BibitemOpen
  \bibfield  {author} {\bibinfo {author} {\bibfnamefont {M.~R.}\ \bibnamefont
  {Wall}}\ and\ \bibinfo {author} {\bibfnamefont {D.}~\bibnamefont
  {Neuhauser}},\ }\bibfield  {title} {\bibinfo {title} {Extraction, through
  filter-diagonalization, of general quantum eigenvalues or classical normal
  mode frequencies from a small number of residues or a short-time segment of a
  signal. i. theory and application to a quantum-dynamics model},\ }\href@noop
  {} {\bibfield  {journal} {\bibinfo  {journal} {J. Chem. Phys.}\ }\textbf
  {\bibinfo {volume} {102}},\ \bibinfo {pages} {8011} (\bibinfo {year}
  {1995})}\BibitemShut {NoStop}%
\bibitem [{\citenamefont {Mandelshtam}\ and\ \citenamefont
  {Taylor}(1997)}]{mandelshtam1997low}%
  \BibitemOpen
  \bibfield  {author} {\bibinfo {author} {\bibfnamefont {V.~A.}\ \bibnamefont
  {Mandelshtam}}\ and\ \bibinfo {author} {\bibfnamefont {H.~S.}\ \bibnamefont
  {Taylor}},\ }\bibfield  {title} {\bibinfo {title} {A low-storage filter
  diagonalization method for quantum eigenenergy calculation or for spectral
  analysis of time signals},\ }\href@noop {} {\bibfield  {journal} {\bibinfo
  {journal} {J. Chem. Phys.}\ }\textbf {\bibinfo {volume} {106}},\ \bibinfo
  {pages} {5085} (\bibinfo {year} {1997})}\BibitemShut {NoStop}%
\bibitem [{\citenamefont {Francis}\ \emph {et~al.}(2022)\citenamefont
  {Francis}, \citenamefont {Agrawal}, \citenamefont {Howard}, \citenamefont
  {K{\"o}kc{\"u}},\ and\ \citenamefont {Kemper}}]{francis2022subspace}%
  \BibitemOpen
  \bibfield  {author} {\bibinfo {author} {\bibfnamefont {A.}~\bibnamefont
  {Francis}}, \bibinfo {author} {\bibfnamefont {A.~A.}\ \bibnamefont
  {Agrawal}}, \bibinfo {author} {\bibfnamefont {J.~H.}\ \bibnamefont {Howard}},
  \bibinfo {author} {\bibfnamefont {E.}~\bibnamefont {K{\"o}kc{\"u}}},\ and\
  \bibinfo {author} {\bibfnamefont {A.}~\bibnamefont {Kemper}},\ }\bibfield
  {title} {\bibinfo {title} {Subspace diagonalization on quantum computers
  using eigenvector continuation},\ }\href@noop {} {\bibfield  {journal}
  {\bibinfo  {journal} {arXiv preprint arXiv:2209.10571}\ } (\bibinfo {year}
  {2022})}\BibitemShut {NoStop}%
\bibitem [{\citenamefont {Tkachenko}\ \emph {et~al.}(2022)\citenamefont
  {Tkachenko}, \citenamefont {Zhang}, \citenamefont {Cincio}, \citenamefont
  {Boldyrev}, \citenamefont {Tretiak},\ and\ \citenamefont
  {Dub}}]{tkachenko2022quantum}%
  \BibitemOpen
  \bibfield  {author} {\bibinfo {author} {\bibfnamefont {N.~V.}\ \bibnamefont
  {Tkachenko}}, \bibinfo {author} {\bibfnamefont {Y.}~\bibnamefont {Zhang}},
  \bibinfo {author} {\bibfnamefont {L.}~\bibnamefont {Cincio}}, \bibinfo
  {author} {\bibfnamefont {A.~I.}\ \bibnamefont {Boldyrev}}, \bibinfo {author}
  {\bibfnamefont {S.}~\bibnamefont {Tretiak}},\ and\ \bibinfo {author}
  {\bibfnamefont {P.~A.}\ \bibnamefont {Dub}},\ }\bibfield  {title} {\bibinfo
  {title} {Quantum davidson algorithm for excited states},\ }\href@noop {}
  {\bibfield  {journal} {\bibinfo  {journal} {arXiv preprint arXiv:2204.10741}\
  } (\bibinfo {year} {2022})}\BibitemShut {NoStop}%
\bibitem [{\citenamefont {Epperly}\ \emph {et~al.}(2022)\citenamefont
  {Epperly}, \citenamefont {Lin},\ and\ \citenamefont
  {Nakatsukasa}}]{epperly2022theory}%
  \BibitemOpen
  \bibfield  {author} {\bibinfo {author} {\bibfnamefont {E.~N.}\ \bibnamefont
  {Epperly}}, \bibinfo {author} {\bibfnamefont {L.}~\bibnamefont {Lin}},\ and\
  \bibinfo {author} {\bibfnamefont {Y.}~\bibnamefont {Nakatsukasa}},\
  }\bibfield  {title} {\bibinfo {title} {A theory of quantum subspace
  diagonalization},\ }\href@noop {} {\bibfield  {journal} {\bibinfo  {journal}
  {SIAM J. Matrix Anal. Appl.}\ }\textbf {\bibinfo {volume} {43}},\ \bibinfo
  {pages} {1263} (\bibinfo {year} {2022})}\BibitemShut {NoStop}%
\bibitem [{\citenamefont {Chen}\ \emph {et~al.}(2021)\citenamefont {Chen},
  \citenamefont {Huang}, \citenamefont {Kueng},\ and\ \citenamefont
  {Tropp}}]{chem:2021concentraion}%
  \BibitemOpen
  \bibfield  {author} {\bibinfo {author} {\bibfnamefont {C.-F.}\ \bibnamefont
  {Chen}}, \bibinfo {author} {\bibfnamefont {H.-Y.}\ \bibnamefont {Huang}},
  \bibinfo {author} {\bibfnamefont {R.}~\bibnamefont {Kueng}},\ and\ \bibinfo
  {author} {\bibfnamefont {J.~A.}\ \bibnamefont {Tropp}},\ }\bibfield  {title}
  {\bibinfo {title} {Concentration for random product formulas},\ }\href
  {https://doi.org/10.1103/PRXQuantum.2.040305} {\bibfield  {journal} {\bibinfo
   {journal} {PRX Quantum}\ }\textbf {\bibinfo {volume} {2}},\ \bibinfo {pages}
  {040305} (\bibinfo {year} {2021})}\BibitemShut {NoStop}%
\bibitem [{\citenamefont {Cho}\ \emph {et~al.}(2022)\citenamefont {Cho},
  \citenamefont {Berry},\ and\ \citenamefont {Hsieh}}]{cho2022doubling}%
  \BibitemOpen
  \bibfield  {author} {\bibinfo {author} {\bibfnamefont {C.~H.}\ \bibnamefont
  {Cho}}, \bibinfo {author} {\bibfnamefont {D.~W.}\ \bibnamefont {Berry}},\
  and\ \bibinfo {author} {\bibfnamefont {M.-H.}\ \bibnamefont {Hsieh}},\
  }\bibfield  {title} {\bibinfo {title} {Doubling the order of approximation
  via the randomized product formula},\ }\href@noop {} {\bibfield  {journal}
  {\bibinfo  {journal} {arXiv preprint arXiv:2210.11281}\ } (\bibinfo {year}
  {2022})}\BibitemShut {NoStop}%
\bibitem [{\citenamefont {Preskill}(2018)}]{preskill2018quantum}%
  \BibitemOpen
  \bibfield  {author} {\bibinfo {author} {\bibfnamefont {J.}~\bibnamefont
  {Preskill}},\ }\bibfield  {title} {\bibinfo {title} {Quantum computing in the
  nisq era and beyond},\ }\href@noop {} {\bibfield  {journal} {\bibinfo
  {journal} {Quantum}\ }\textbf {\bibinfo {volume} {2}},\ \bibinfo {pages} {79}
  (\bibinfo {year} {2018})}\BibitemShut {NoStop}%
\bibitem [{\citenamefont {Kirby}\ \emph {et~al.}(2022)\citenamefont {Kirby},
  \citenamefont {Motta},\ and\ \citenamefont {Mezzacapo}}]{kirby2022exact}%
  \BibitemOpen
  \bibfield  {author} {\bibinfo {author} {\bibfnamefont {W.}~\bibnamefont
  {Kirby}}, \bibinfo {author} {\bibfnamefont {M.}~\bibnamefont {Motta}},\ and\
  \bibinfo {author} {\bibfnamefont {A.}~\bibnamefont {Mezzacapo}},\ }\bibfield
  {title} {\bibinfo {title} {Exact and efficient lanczos method on a quantum
  computer},\ }\href@noop {} {\bibfield  {journal} {\bibinfo  {journal} {arXiv
  preprint arXiv:2208.00567}\ } (\bibinfo {year} {2022})}\BibitemShut {NoStop}%
\bibitem [{\citenamefont {Wang}\ \emph {et~al.}(2022)\citenamefont {Wang},
  \citenamefont {Stilck-Fran{\c{c}}a}, \citenamefont {Zhang}, \citenamefont
  {Zhu},\ and\ \citenamefont {Johnson}}]{wang2022quantum}%
  \BibitemOpen
  \bibfield  {author} {\bibinfo {author} {\bibfnamefont {G.}~\bibnamefont
  {Wang}}, \bibinfo {author} {\bibfnamefont {D.}~\bibnamefont
  {Stilck-Fran{\c{c}}a}}, \bibinfo {author} {\bibfnamefont {R.}~\bibnamefont
  {Zhang}}, \bibinfo {author} {\bibfnamefont {S.}~\bibnamefont {Zhu}},\ and\
  \bibinfo {author} {\bibfnamefont {P.~D.}\ \bibnamefont {Johnson}},\
  }\bibfield  {title} {\bibinfo {title} {Quantum algorithm for ground state
  energy estimation using circuit depth with exponentially improved dependence
  on precision},\ }\href@noop {} {\bibfield  {journal} {\bibinfo  {journal}
  {arXiv preprint arXiv:2209.06811}\ } (\bibinfo {year} {2022})}\BibitemShut
  {NoStop}%
\bibitem [{\citenamefont {Golub}\ and\ \citenamefont
  {Van~Loan}(2013)}]{golub2013matrix}%
  \BibitemOpen
  \bibfield  {author} {\bibinfo {author} {\bibfnamefont {G.~H.}\ \bibnamefont
  {Golub}}\ and\ \bibinfo {author} {\bibfnamefont {C.~F.}\ \bibnamefont
  {Van~Loan}},\ }\href@noop {} {\emph {\bibinfo {title} {Matrix
  computations}}}\ (\bibinfo  {publisher} {JHU press},\ \bibinfo {year}
  {2013})\BibitemShut {NoStop}%
\bibitem [{\citenamefont {Childs}\ \emph {et~al.}(2021)\citenamefont {Childs},
  \citenamefont {Su}, \citenamefont {Tran}, \citenamefont {Wiebe},\ and\
  \citenamefont {Zhu}}]{childs2021theory}%
  \BibitemOpen
  \bibfield  {author} {\bibinfo {author} {\bibfnamefont {A.~M.}\ \bibnamefont
  {Childs}}, \bibinfo {author} {\bibfnamefont {Y.}~\bibnamefont {Su}}, \bibinfo
  {author} {\bibfnamefont {M.~C.}\ \bibnamefont {Tran}}, \bibinfo {author}
  {\bibfnamefont {N.}~\bibnamefont {Wiebe}},\ and\ \bibinfo {author}
  {\bibfnamefont {S.}~\bibnamefont {Zhu}},\ }\bibfield  {title} {\bibinfo
  {title} {Theory of trotter error with commutator scaling},\ }\href@noop {}
  {\bibfield  {journal} {\bibinfo  {journal} {Phys. Rev. X}\ }\textbf {\bibinfo
  {volume} {11}},\ \bibinfo {pages} {011020} (\bibinfo {year}
  {2021})}\BibitemShut {NoStop}%
\bibitem [{\citenamefont {K{\'a}llay}(2014)}]{kallay2014systematic}%
  \BibitemOpen
  \bibfield  {author} {\bibinfo {author} {\bibfnamefont {M.}~\bibnamefont
  {K{\'a}llay}},\ }\bibfield  {title} {\bibinfo {title} {A systematic way for
  the cost reduction of density fitting methods},\ }\href@noop {} {\bibfield
  {journal} {\bibinfo  {journal} {J. Chem. Phys.}\ }\textbf {\bibinfo {volume}
  {141}},\ \bibinfo {pages} {244113} (\bibinfo {year} {2014})}\BibitemShut
  {NoStop}%
\bibitem [{\citenamefont {Koridon}\ \emph {et~al.}(2021)\citenamefont
  {Koridon}, \citenamefont {Yalouz}, \citenamefont {Senjean}, \citenamefont
  {Buda}, \citenamefont {O'Brien},\ and\ \citenamefont
  {Visscher}}]{koridon2021orbital}%
  \BibitemOpen
  \bibfield  {author} {\bibinfo {author} {\bibfnamefont {E.}~\bibnamefont
  {Koridon}}, \bibinfo {author} {\bibfnamefont {S.}~\bibnamefont {Yalouz}},
  \bibinfo {author} {\bibfnamefont {B.}~\bibnamefont {Senjean}}, \bibinfo
  {author} {\bibfnamefont {F.}~\bibnamefont {Buda}}, \bibinfo {author}
  {\bibfnamefont {T.~E.}\ \bibnamefont {O'Brien}},\ and\ \bibinfo {author}
  {\bibfnamefont {L.}~\bibnamefont {Visscher}},\ }\bibfield  {title} {\bibinfo
  {title} {Orbital transformations to reduce the 1-norm of the electronic
  structure hamiltonian for quantum computing applications},\ }\href@noop {}
  {\bibfield  {journal} {\bibinfo  {journal} {Phys. Rev. Res.}\ }\textbf
  {\bibinfo {volume} {3}},\ \bibinfo {pages} {033127} (\bibinfo {year}
  {2021})}\BibitemShut {NoStop}%
\bibitem [{\citenamefont {Hagan}\ and\ \citenamefont
  {Wiebe}(2022)}]{hagan2022composite}%
  \BibitemOpen
  \bibfield  {author} {\bibinfo {author} {\bibfnamefont {M.}~\bibnamefont
  {Hagan}}\ and\ \bibinfo {author} {\bibfnamefont {N.}~\bibnamefont {Wiebe}},\
  }\bibfield  {title} {\bibinfo {title} {Composite quantum simulations},\
  }\href@noop {} {\bibfield  {journal} {\bibinfo  {journal} {arXiv preprint
  arXiv:2206.06409}\ } (\bibinfo {year} {2022})}\BibitemShut {NoStop}%
\bibitem [{\citenamefont {Rajput}\ \emph {et~al.}(2022)\citenamefont {Rajput},
  \citenamefont {Roggero},\ and\ \citenamefont {Wiebe}}]{rajput2022hybridized}%
  \BibitemOpen
  \bibfield  {author} {\bibinfo {author} {\bibfnamefont {A.}~\bibnamefont
  {Rajput}}, \bibinfo {author} {\bibfnamefont {A.}~\bibnamefont {Roggero}},\
  and\ \bibinfo {author} {\bibfnamefont {N.}~\bibnamefont {Wiebe}},\ }\bibfield
   {title} {\bibinfo {title} {Hybridized methods for quantum simulation in the
  interaction picture},\ }\href@noop {} {\bibfield  {journal} {\bibinfo
  {journal} {Quantum}\ }\textbf {\bibinfo {volume} {6}},\ \bibinfo {pages}
  {780} (\bibinfo {year} {2022})}\BibitemShut {NoStop}%
\bibitem [{\citenamefont {Sayfutyarova}\ \emph {et~al.}(2017)\citenamefont
  {Sayfutyarova}, \citenamefont {Sun}, \citenamefont {Chan},\ and\
  \citenamefont {Knizia}}]{Sayfutyarova_2017}%
  \BibitemOpen
  \bibfield  {author} {\bibinfo {author} {\bibfnamefont {E.~R.}\ \bibnamefont
  {Sayfutyarova}}, \bibinfo {author} {\bibfnamefont {Q.}~\bibnamefont {Sun}},
  \bibinfo {author} {\bibfnamefont {G.~K.-L.}\ \bibnamefont {Chan}},\ and\
  \bibinfo {author} {\bibfnamefont {G.}~\bibnamefont {Knizia}},\ }\bibfield
  {title} {\bibinfo {title} {Automated construction of molecular active spaces
  from atomic valence orbitals},\ }\href
  {https://doi.org/10.1021/acs.jctc.7b00128} {\bibfield  {journal} {\bibinfo
  {journal} {J. Chem. Theory Comput.}\ }\textbf {\bibinfo {volume} {13}},\
  \bibinfo {pages} {4063} (\bibinfo {year} {2017})}\BibitemShut {NoStop}%
\bibitem [{\citenamefont {Sun}\ \emph {et~al.}(2020)\citenamefont {Sun},
  \citenamefont {Zhang}, \citenamefont {Banerjee}, \citenamefont {Bao},
  \citenamefont {Barbry}, \citenamefont {Blunt}, \citenamefont {Bogdanov},
  \citenamefont {Booth}, \citenamefont {Chen}, \citenamefont {Cui} \emph
  {et~al.}}]{sun2020recent}%
  \BibitemOpen
  \bibfield  {author} {\bibinfo {author} {\bibfnamefont {Q.}~\bibnamefont
  {Sun}}, \bibinfo {author} {\bibfnamefont {X.}~\bibnamefont {Zhang}}, \bibinfo
  {author} {\bibfnamefont {S.}~\bibnamefont {Banerjee}}, \bibinfo {author}
  {\bibfnamefont {P.}~\bibnamefont {Bao}}, \bibinfo {author} {\bibfnamefont
  {M.}~\bibnamefont {Barbry}}, \bibinfo {author} {\bibfnamefont {N.~S.}\
  \bibnamefont {Blunt}}, \bibinfo {author} {\bibfnamefont {N.~A.}\ \bibnamefont
  {Bogdanov}}, \bibinfo {author} {\bibfnamefont {G.~H.}\ \bibnamefont {Booth}},
  \bibinfo {author} {\bibfnamefont {J.}~\bibnamefont {Chen}}, \bibinfo {author}
  {\bibfnamefont {Z.-H.}\ \bibnamefont {Cui}}, \emph {et~al.},\ }\bibfield
  {title} {\bibinfo {title} {Recent developments in the pyscf program
  package},\ }\href@noop {} {\bibfield  {journal} {\bibinfo  {journal} {J.
  Chem. Phys.}\ }\textbf {\bibinfo {volume} {153}},\ \bibinfo {pages} {024109}
  (\bibinfo {year} {2020})}\BibitemShut {NoStop}%
\bibitem [{\citenamefont {Peruzzo}\ \emph {et~al.}(2014)\citenamefont
  {Peruzzo}, \citenamefont {McClean}, \citenamefont {Shadbolt}, \citenamefont
  {Yung}, \citenamefont {Zhou}, \citenamefont {Love}, \citenamefont
  {Aspuru-Guzik},\ and\ \citenamefont {O'brien}}]{peruzzo2014variational}%
  \BibitemOpen
  \bibfield  {author} {\bibinfo {author} {\bibfnamefont {A.}~\bibnamefont
  {Peruzzo}}, \bibinfo {author} {\bibfnamefont {J.}~\bibnamefont {McClean}},
  \bibinfo {author} {\bibfnamefont {P.}~\bibnamefont {Shadbolt}}, \bibinfo
  {author} {\bibfnamefont {M.-H.}\ \bibnamefont {Yung}}, \bibinfo {author}
  {\bibfnamefont {X.-Q.}\ \bibnamefont {Zhou}}, \bibinfo {author}
  {\bibfnamefont {P.~J.}\ \bibnamefont {Love}}, \bibinfo {author}
  {\bibfnamefont {A.}~\bibnamefont {Aspuru-Guzik}},\ and\ \bibinfo {author}
  {\bibfnamefont {J.~L.}\ \bibnamefont {O'brien}},\ }\bibfield  {title}
  {\bibinfo {title} {A variational eigenvalue solver on a photonic quantum
  processor},\ }\href@noop {} {\bibfield  {journal} {\bibinfo  {journal} {Nat.
  Commun.}\ }\textbf {\bibinfo {volume} {5}},\ \bibinfo {pages} {1} (\bibinfo
  {year} {2014})}\BibitemShut {NoStop}%
\bibitem [{\citenamefont {Yung}\ \emph {et~al.}(2014)\citenamefont {Yung},
  \citenamefont {Casanova}, \citenamefont {Mezzacapo}, \citenamefont {Mcclean},
  \citenamefont {Lamata}, \citenamefont {Aspuru-Guzik},\ and\ \citenamefont
  {Solano}}]{yung2014transistor}%
  \BibitemOpen
  \bibfield  {author} {\bibinfo {author} {\bibfnamefont {M.-H.}\ \bibnamefont
  {Yung}}, \bibinfo {author} {\bibfnamefont {J.}~\bibnamefont {Casanova}},
  \bibinfo {author} {\bibfnamefont {A.}~\bibnamefont {Mezzacapo}}, \bibinfo
  {author} {\bibfnamefont {J.}~\bibnamefont {Mcclean}}, \bibinfo {author}
  {\bibfnamefont {L.}~\bibnamefont {Lamata}}, \bibinfo {author} {\bibfnamefont
  {A.}~\bibnamefont {Aspuru-Guzik}},\ and\ \bibinfo {author} {\bibfnamefont
  {E.}~\bibnamefont {Solano}},\ }\bibfield  {title} {\bibinfo {title} {From
  transistor to trapped-ion computers for quantum chemistry},\ }\href@noop {}
  {\bibfield  {journal} {\bibinfo  {journal} {Sci. Rep.}\ }\textbf {\bibinfo
  {volume} {4}},\ \bibinfo {pages} {1} (\bibinfo {year} {2014})}\BibitemShut
  {NoStop}%
\bibitem [{\citenamefont {McClean}\ \emph {et~al.}(2016)\citenamefont
  {McClean}, \citenamefont {Romero}, \citenamefont {Babbush},\ and\
  \citenamefont {Aspuru-Guzik}}]{mcclean2016theory}%
  \BibitemOpen
  \bibfield  {author} {\bibinfo {author} {\bibfnamefont {J.~R.}\ \bibnamefont
  {McClean}}, \bibinfo {author} {\bibfnamefont {J.}~\bibnamefont {Romero}},
  \bibinfo {author} {\bibfnamefont {R.}~\bibnamefont {Babbush}},\ and\ \bibinfo
  {author} {\bibfnamefont {A.}~\bibnamefont {Aspuru-Guzik}},\ }\bibfield
  {title} {\bibinfo {title} {The theory of variational hybrid quantum-classical
  algorithms},\ }\href@noop {} {\bibfield  {journal} {\bibinfo  {journal} {New
  J. Phys.}\ }\textbf {\bibinfo {volume} {18}},\ \bibinfo {pages} {023023}
  (\bibinfo {year} {2016})}\BibitemShut {NoStop}%
\bibitem [{\citenamefont {Huang}\ \emph {et~al.}(2020)\citenamefont {Huang},
  \citenamefont {Kueng},\ and\ \citenamefont {Preskill}}]{huang2020predicting}%
  \BibitemOpen
  \bibfield  {author} {\bibinfo {author} {\bibfnamefont {H.-Y.}\ \bibnamefont
  {Huang}}, \bibinfo {author} {\bibfnamefont {R.}~\bibnamefont {Kueng}},\ and\
  \bibinfo {author} {\bibfnamefont {J.}~\bibnamefont {Preskill}},\ }\bibfield
  {title} {\bibinfo {title} {Predicting many properties of a quantum system
  from very few measurements},\ }\href@noop {} {\bibfield  {journal} {\bibinfo
  {journal} {Nat. Phys.}\ }\textbf {\bibinfo {volume} {16}},\ \bibinfo {pages}
  {1050} (\bibinfo {year} {2020})}\BibitemShut {NoStop}%
\bibitem [{\citenamefont {Wecker}\ \emph {et~al.}(2015)\citenamefont {Wecker},
  \citenamefont {Hastings}, \citenamefont {Wiebe}, \citenamefont {Clark},
  \citenamefont {Nayak},\ and\ \citenamefont {Troyer}}]{wecker2015solving}%
  \BibitemOpen
  \bibfield  {author} {\bibinfo {author} {\bibfnamefont {D.}~\bibnamefont
  {Wecker}}, \bibinfo {author} {\bibfnamefont {M.~B.}\ \bibnamefont
  {Hastings}}, \bibinfo {author} {\bibfnamefont {N.}~\bibnamefont {Wiebe}},
  \bibinfo {author} {\bibfnamefont {B.~K.}\ \bibnamefont {Clark}}, \bibinfo
  {author} {\bibfnamefont {C.}~\bibnamefont {Nayak}},\ and\ \bibinfo {author}
  {\bibfnamefont {M.}~\bibnamefont {Troyer}},\ }\bibfield  {title} {\bibinfo
  {title} {Solving strongly correlated electron models on a quantum computer},\
  }\href {https://journals.aps.org/pra/abstract/10.1103/PhysRevA.92.062318}
  {\bibfield  {journal} {\bibinfo  {journal} {Phys. Rev. A}\ }\textbf {\bibinfo
  {volume} {92}},\ \bibinfo {pages} {062318} (\bibinfo {year}
  {2015})}\BibitemShut {NoStop}%
\bibitem [{\citenamefont {McLachlan}\ and\ \citenamefont
  {Peel}(2000)}]{McLachlan2000Finite}%
  \BibitemOpen
  \bibfield  {author} {\bibinfo {author} {\bibfnamefont {G.~J.}\ \bibnamefont
  {McLachlan}}\ and\ \bibinfo {author} {\bibfnamefont {D.}~\bibnamefont
  {Peel}},\ }\href@noop {} {\emph {\bibinfo {title} {Finite Mixture Models}}}\
  (\bibinfo  {publisher} {Wiley},\ \bibinfo {year} {2000})\BibitemShut
  {NoStop}%
\end{thebibliography}%

\end{document}